\documentclass[apj]{emulateapj}

\usepackage{natbib}
\usepackage{graphicx}
\usepackage{xspace}
\usepackage{url}
\usepackage{hyperref}
\usepackage{multirow}
\newcommand{\sout}[1]{}

\def\btheta{\mathbf{\theta}}

\def\ie{{\it i.e.}}

\renewcommand{\d}{\mathrm{d}}

\newcommand{\Mpc}{\mathrm{Mpc}}
\newcommand{\Msol}{\mathrm{M}_\odot} 
\newcommand{\gamt}{\langle \gamma_t \rangle}
\renewcommand{\b}{\langle b \rangle}
\renewcommand{\r}{\langle r \rangle}
\newcommand{\nmx}{\langle NM_\mathrm{x} \rangle}

\newcommand{\map}{\langle M^2_\mathrm{ap}(\theta)\rangle}
\newcommand{\mpe}{\langle M^2_\perp (\theta)\rangle}
\newcommand{\nap}{\langle \mathcal{N}^2(\theta)\rangle}
\newcommand{\nmap}{\langle
\mathcal{N}(\theta)M_\mathrm{ap}(\theta) \rangle}

\begin{document}

\title{COSMOS: Stochastic bias from measurements of weak lensing and
galaxy clustering}
\author{Eric Jullo\altaffilmark{1,2}}
\author{Jason Rhodes\altaffilmark{1}}
\author{Alina Kiessling\altaffilmark{3}}
\author{James E. Taylor\altaffilmark{4}}
\author{Richard Massey\altaffilmark{3}}
\author{Joel Berge\altaffilmark{5}}
\author{Carlo Schimd\altaffilmark{2}}
\author{Jean-Paul Kneib\altaffilmark{2}}
\author{Nick Scoville\altaffilmark{6}}

\altaffiltext{1}{Propulsion Laboratory, California Institute of
Technology, Pasadena, CA 91109, USA}
\altaffiltext{2}{Laboratoire d'Astrophysique de Marseille,
Universit\'e de Provence, CNRS, 13388 Marseille CEDEX 13, France}
\altaffiltext{3}{Institute for Astronomy, Blackford Hill, Edinburgh EH9 3HJ
UK}
\altaffiltext{4}{Department of Physics and Astronomy, University of Waterloo,
200 University Avenue West, Waterloo, Ontario, Canada N2L}
\altaffiltext{5}{Institute of Astronomy, Department of Physics, ETH Zurich,
CH-8093, Switzerland}
\altaffiltext{6}{California Institute of Technology, MC 105-24, 1200 East California
Boulevard, Pasadena, CA 91125, USA}
\email{eric.jullo@oamp.fr}

\begin{abstract}

In the theory of structure formation, galaxies are biased tracers of the underlying matter density field.  The statistical relation between galaxy and matter density field is commonly referred as galaxy bias. In this paper, we test the linear bias model with  weak-lensing and galaxy clustering measurements in
the 2 square degrees COSMOS field \citep{scoville2007}.  We estimate the bias of galaxies between redshifts $z=0.2$ and $z=1$, and over correlation scales 
 between $R=0.2\ h^{-1}\ \Mpc$ and $R=15\ h^{-1}\ \Mpc$. We focus on three galaxy samples, selected in flux (simultaneous cuts $I_{814W} <
26.5$ and $K_s < 24$), and in
stellar-mass ($10^9 < M_* < 10^{10}\ h^{-2}\ \Msol$ and $10^{10} < M_*
< 10^{11}\ h^{-2}\ \Msol$).  At scales $R > 2\ h^{-1}\Mpc$, our
measurements support a model of bias increasing with redshift. The
Tinker et al. (2010) fitting function provides a good fit to the data.
We find the best fit mass of the galaxy halos to be
$\log(M_{200}/h^{-1}\Msol) = 11.7^{+0.6}_{-1.3}$ and
$\log(M_{200}/h^{-1}\Msol) = 12.4^{+0.2}_{-2.9}$ respectively for the
low and high stellar-mass samples. In the halo model framework, bias is scale-dependent with a change of slope at the transition scale between the one and the two halo terms. We detect a scale-dependence of
bias with a turn-down at scale $R=2.3\pm1.5 h^{-1}\Mpc$, in agreement with previous galaxy clustering studies. 
We find no significant amount of stochasticity,
suggesting that a linear bias model is sufficient to describe our
data. We use N-body simulations to quantify both the amount of cosmic
variance and systematic errors in the measurement.

\end{abstract}


\section{Introduction}

In the theory of structure formation, galaxies form in dark matter overdensities. However, the distribution of galaxies does not perfectly match the underlying matter distribution, and the relation between the two is called galaxy bias. 

The galaxy bias relation $\delta_g = f(\delta)$ relates the matter
density contrast $\delta$ to the galaxy density contrast $\delta_g$.
The density contrast $\delta$ is defined as the ratio between the
local and the mean densities $\delta \equiv \rho / \bar{\rho} -1$.  To
first order, the galaxy bias relation $\delta_g = b \delta$ is
 parametrized by the bias parameter $b$. However the stochasticity in the physical processes of galaxy formation might introduce some stochasticity in the 
galaxy bias relation. The amount of stochasticity can be measured with the correlation coefficient $r$ defined as the ratio between the covariance and the variance of the density
contrasts $\delta$ and $\delta_g$,  $r = \langle \delta_g \delta
\rangle / \sqrt{\langle \delta_g^2 \rangle \langle \delta^2 \rangle}$.
A correlation coefficient $r=1$ means the bias relation is perfectly
linear, whereas $0 < r < 1$ suggests stochasticity  and/or non-linearity \citep{dekel1999}. We do not expect
the bias relation to be anti-correlated with $r<0$, nor $b<0$ (\ie\ galaxy clustering decreases when matter clustering increases).

Bias varies for different galaxy populations. Galaxy biasing is known to be
larger for luminous, red, and high-redshift galaxies
\citep{marinoni2005,meneux2006,coil2008,zehavi2010}. Bias also varies
with scale, but differently for different galaxy types and luminosity
\citep{cresswell2009,ouchi2005,mccracken2010}.  In the halo-model
framework, the scale-dependence of bias is explained as the transition
from a small-scale regime where galaxies are in the same halo of
dark matter to a large-scale regime where galaxies are in two
different halos \citep[see e.g.][for a recent discussion]{zheng2009}.
The scale-dependence of bias therefore tells us about the dark matter
halo properties around galaxies.

\citet{tegmark1999} first measured bias stochasticity in the Las Campanas Redshift
Survey. At the same time, simulations started to predict
stochasticity,
as well as
scale and redshift-dependence in the bias relation due
to physical processes
\citep{blanton1999,somerville2001,yoshikawa2001}. In the 2dF Galaxy
Redshift Survey, \citet{wild2005} used the count-in-cell technique to
measure the non-linearity and stochasticity of the bias relation for
red/early and blue/late-type galaxies at redshift $z < 0.114$. For
both categories, they found more stochasticity at scale $R=10$ Mpc
than at scale $R=45$ Mpc. Recent results from simulations also predict
an increasing amount of stochasticity at smaller scales
\citep{baldauf2010}.

The correlation coefficient is at the same time an estimate of stochasticity and non-linearity.
In the zCOSMOS redshift
survey \citep{lilly2009}, \citet{kovac2009} found non-linearity to
contribute less than $\sim 0.2$\% to the bias relation for
luminosity-selected galaxies ($M_B < -20 -z$). Therefore in the following, we will assume that the correlation coefficient $r$  mostly quantifies bias stochasticity. With respect to the bias parameter $b$, \citet{kovac2009} found the bias relation to be
scale-independent between scales $R=8\ h^{-1}\ \Mpc$ and $R=10\
h^{-1}\ \Mpc$, and redshift-dependent between redshifts $z=0.4$ and
$z=1$. 

In this work, we measure galaxy bias with a technique proposed by \citet{schneider1998b} based on galaxy clustering and weak lensing.  In short, weak lensing is used to derive matter clustering from the shape of background galaxies, which is then compared to the galaxy clustering to infer the bias.   \citet{vanwaerbeke1999}
emphasized the potential of this technique with analytical
calculations for surveys with various depths and areas, and
\citet{hoekstra2001,hoekstra2002} applied it to
the 50.5 deg$^2$ of the RCS and VIRMOS-DESCART surveys.  They measured
the linear bias parameter $b$ and the correlation coefficient $r$ for a
flux-limited galaxy sample ($19.5 < R_C < 21$) at redshift $\bar{z}
\simeq 0.35$, and on scales between $R= 0.2$ and $9.3\ h^{-1}_{50}\
\Mpc$.
They found strong evidence that both $b$ and $r$ change with
scale, and $r\sim 0.57$ at $1\ h^{-1}_{50}\ \Mpc$, which suggests a
significant degree of stochasticity and/or non-linearity in their
sample of galaxies.  \citet{simonp2007} applied the same technique  to the
15 deg$^2$ of the GaBoDS survey for three flux-limited galaxy samples
in the R-band, at redshift $\bar{z}=$0.35, 0.47 and 0.61. They also found
bias to be scale dependent, with an increasing amount of bias at small
scales. In addition, they found the bias parameters $b$ and $r$ to be
redshift-independent within statistical uncertainties, with
$r\sim0.6$.  Finally, \citet{sheldon2004} performed a comparable study
in the 3800 deg$^2$ of the SDSS survey, with volume- and
magnitude-selected galaxies ($0.1 < z < 0.174$ and $-23.0 < M_r - 5
\log h < -21.5$).  In agreement with
\citet{hoekstra2002}, they found the ratio $b/r$ to be scale-independent
between 0.4 and 6.7 $h^{-1}$\ Mpc, with $(b/r) = (1.3\pm0.2)(\Omega_m
/ 0.27)$ for a flux-limited galaxy sample.  In fact, the \citet{hoekstra2002} results suggested
that the respective scale-dependence of $b$ and $r$ conspire to produce a scale-independent ratio $b/r$. In summary, the consistent
value of $r<1$ found in all these studies suggests some degree of
stochasticity in the investigated galaxy samples.

In this work, we study the bias of galaxies in the 2 square degrees of the COSMOS field. We test the linear bias model, and its dependence on scale and redshift. For two galaxy samples selected in stellar mass and one sample selected in flux, we measure bias on scales between $0.2\ h^{-1}\ \Mpc$ and
$15\ h^{-1}\ \Mpc$, and redshifts between $z=0.2$ and $z=1$. Thus, we extend the previous works on bias both in scale, redshift and stellar
mass. COSMOS is ideal to perform this study
because accurate stellar mass have been measured up to redshift $z\sim1.2$ \citep{bundy2010}, and photometric redshifts up to redshift  $z\sim 4$   \citep{ilbert2009}. 
In addition, with the high resolution images from the 
 the Advanced Camera for Survey aboard the Hubble Space Telescope
(ACS/HST), we can precisely measure  the shape of more than
200,000 galaxies. Measuring the shape of galaxies at redshift $z> 2$ is crucial for weak lensing to allow accurate determination of the matter density up to redshift $z\sim1$. 
 In section~\ref{sec:cosmos}, we present the COSMOS
survey, our lensing catalog and our selection of foreground galaxies
for which we want to measure the bias. In section~\ref{sec:formalism},
we describe the lensing formalism used in this paper. In
section~\ref{sec:simulations}, we present a set of simulations we use
to quantify the impact of cosmic variance.  In
section~\ref{sec:results}, we show our results, and we go on to discuss some possible
systematic errors in section~\ref{sec:discussion}. We conclude in
section~\ref{sec:conclusion}.

In this paper, we assume the WMAP7 \citep{komatsu2010} cosmological
parameters $\Omega_m = 0.272$, $\Omega_\Lambda=0.728$, $\Omega_b =
0.0449$, $\sigma_8 = 0.809$, $n=0.963$, $\omega_0 = -1$ and $h=0.71$.

\section{DATA}
\label{sec:cosmos}

\subsection{General properties}

The COSMOS field is a 1.64 deg$^2$ patch of sky close to the
equator ($\alpha$=10\fh00\fm28\fs, $\delta$=+2\fdg12\farcm21\farcs).
It was imaged with ACS/HST during cycles 12-13
\citep{scoville2007,koekemoer2007}.  An intensive follow-up from UV to
IR was conducted by several ground-based and space-based telescopes.
In this study, we take advantage of the spatial resolution of the ACS
imaging to measure the shape of the galaxies, and of the multi-filter
ground-based imaging to derive accurate photometric redshifts \citep{ilbert2009}.

\subsection{Lensing catalog}


We use an updated version of the  lensing catalog presented in
\citet{leauthaud2007}. In this new catalog, raw ACS/f814w images have
been corrected for the Charge Transfer Inefficiency (CTI) effect using
the algorithm described in \citet{massey2010}. Objects are detected
with the ``hot-cold'' technique described in \citet{leauthaud2007},
i.e.  Sextractor \citep{bertin1996} is run twice~: first to detect
extended objects, and then compact objects.  Regions contaminated by
bright stars or satellite trails are masked out. Point sources and
spurious detections are discarded as detailed in
\citep{leauthaud2007}.  The catalog is
complete up to magnitude $\mathrm{MAG\_AUTO}=26.5$ in the ACS/f814w
band (hereafter $I_{814W}$).

Galaxy ellipticities are measured with the RRG algorithm described in
\citet{rhodes2000}, and calibrated with the same shear calibration as
in \citet{leauthaud2007}. In our final source catalog, we make the cuts on
size $d > 1.6$ times the PSF size and S/N $>4.5$ as in \citet{leauthaud2007}.  We also remove galaxies with more than one peak in their photometric 
redshift probability distribution function, or galaxies masked in the Subaru imaging. This way, we obtain an average redshift error $\sigma_z / (1+z) < 0.05$
Our source catalog with good shapes and redshifts contains 210,477
galaxies, which yields a density of 36 galaxies per square arc-minute.
As in \citet{massey2007b}, we assign to every source galaxy a inverse variance weight
$w$ aimed at maximizing the S/N, and calculated from the galaxy apparent magnitude 

\begin{equation}
    \label{eq:weight}
    w = \frac{1}{\sigma_\epsilon^{2} +0.1}\,,
\end{equation}

\noindent where the observational noise (pixellation, shot noise, etc) is well approximated by the following function of the galaxy apparent magnitude in the $I_{814W}$ band

\begin{equation}
    \sigma_\epsilon = 0.32 + 0.0014(\mathrm{MAG\_AUTO} - 20)^3\,.
\end{equation}

\noindent  The second term 0.1 represents the variance of the intrinsic shape noise, which was found to be pretty constant in \citet{leauthaud2007}. Note that the power of 2 was omitted as a typo in  \citet{massey2007b}.

\subsection{Foreground and background galaxy catalogs}

\subsubsection{Catalog cuts by redshift}
\label{sec:catalogs}

In this work, we want to study the evolution of galaxy bias as a function of redshift and scale. Bias is computed from the ratio of galaxy and matter clustering in bins of redshift. To ease the interpretation of the bias results later on, we choose to define foreground galaxy samples with redshift distributions matching the redshift distribution of the matter that most efficiently perturbs the background galaxy shapes through weak lensing. Typically, the matter that lies halfway between us and the  background galaxies, is the one that most efficiently perturbs their shape. We show in Section~\ref{sec:discussion} that a mismatch in redshift can change the estimated bias.
 
We create 3 bins B1, B2 and B3 of background sources.  We set the bin limits to $z=0.3$, 0.8, 1.4 and 4, so that the number of galaxies and signal to noise in each bin is similar.
 We assign a galaxy with redshift $z\pm\sigma_{68\%}$ to a
particular redshift bin with limits $z_\mathrm{low}$ and
$z_\mathrm{high}$ if $z - \sigma_{68\%} > z_\mathrm{low}$
and  $z + \sigma_{68\%} < z_\mathrm{high}$. The values of $z$ and $\sigma_{68\%}$ are estimated from the photometric redshift likelihoods. The number of
galaxies per bin is reported in Table~\ref{tab:bins}.

For each  background bin, we compute the lensing efficiency. The lensing efficiency is a calculation based on the redshift distribution of the background galaxies in comoving coordinates $p_b(w) = p_b(z) \d z / \d w$ and of the cosmological parameters.We use it as a weighting function in the following to project the 3D matter power spectrum along the line of sight. The lensing efficiency  as a function of the comoving distance to the lens is defined as  

\begin{equation}
    \label{eq:lensingefficiency}
    g(w) = \frac{3 \Omega_m H_0^2 f_K(w)}{2 c^2 a(w)} \int_w^{w_h} \d w'\ 
    p_b(w') \frac{f_K(w-w')}{f_K{w'}}
\end{equation}

\noindent where the functions $a(w)$ and $f_K(w)$ are respectively the comoving scale factor and the comoving angular distance.  We find the lensing efficiency curves to peak at redshifts
$z_{B1}=0.22$, $z_{B2}=0.37$ and $z_{B3}=0.51$.

Next, we create 5 bins of foreground galaxies. For the three first bins,  we adjust their centers to match  the peaks of the 3 lensing efficiencies. Then, we adjust their width so that the signal to noise of the galaxy-galaxy cross-correlation measurements is the largest. Too broad bins increase the mixing of angular and physical scales, and too narrow bins increase shot noise. Bins must also be broad enough, so that we can use the Limber approximation to compute the theoretical signal. Indeed, the Limber approximation breaks down at large scales for too narrow redshift bins.  \citet{simon2007b} computed the minimal bin width as a function of scale and redshift, beyond which the Limber approximation becomes inaccurate by more than 10\%. To minimize the impact of this issue, we fix the bin width so that we reach 10\% inaccuracy at 100 arc-minutes, \ie\ twice the largest scale at which we compute the correlation functions. In summary, 
we obtain three foreground bins with limits $0.17 < z_\mathrm{F1} <
0.27$, $0.32 < z_\mathrm{F2} < 0.42$, and $0.47 < z_\mathrm{F3} <
0.57$.  To these 3 foreground bins, we add two extra bins to probe the
bias at higher redshift $0.60 < z_{F4} < 0.75$ and $0.87 <
z_\mathrm{F5} < 1.07$. Finally, we assign the foreground galaxies to their respective bins using 
the same criteria we used for the background sources, and compute  $p_f(z)$ the redshift distribution of the  foreground bins. Again, we will use these redshift distributions as weighting functions to project the 3D matter power spectrum along the line of sight, and compute the different estimators presented below.

\begin{figure*}
    \centering
    \includegraphics[width=\linewidth]{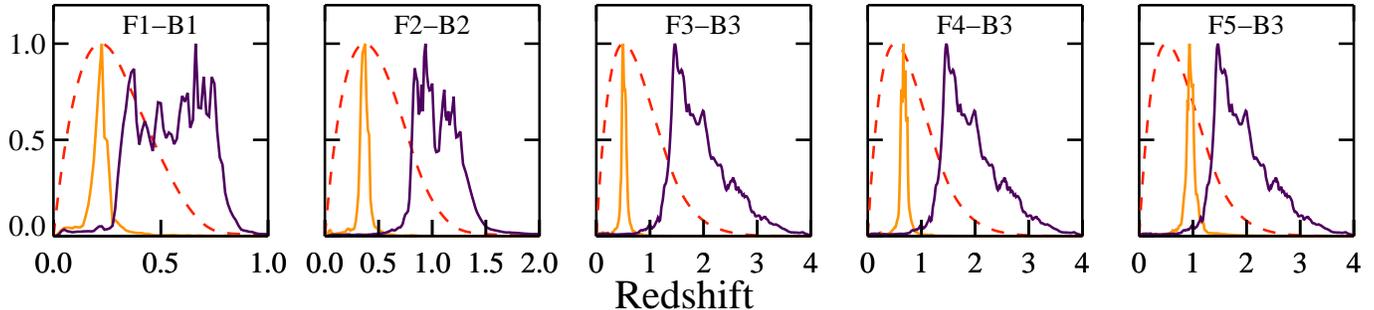}
    \caption{Redshift distributions of the 3 background catalogs
    B1--B3 (purple/dark line), and the five foreground catalogs F1--F5 (yellow/light line).
    The normalized lensing efficiency curves (red dashed lines) give the
      location where the matter most efficiently perturbs the shape of the background galaxies. [{\it See the electronic edition for a color version of this figure.}]}

    \label{fig:lensingEfficiencies}
\end{figure*}

Note here three important points
regarding our analysis. First, we compute the redshift distribution of the galaxies that fall in the foreground and background redshift bins $p_f(z)$ and $p_b(z)$ by summing their redshift posterior probability distribution function (pdf) $p^{(i)}(z)$. The redshift distributions $p_f(z)$ and $p_b(z)$, as well as the lensing efficiency curves are shown in Figure~\ref{fig:lensingEfficiencies}.  Summing the redshift pdfs instead of building histograms of the individual redshift estimates has several advantages~: (i) it helps to deal with skewed pdfs,  (ii) it tells us about the probability of having galaxies outside a redshift bin. In spite of our cuts in redshift, we still find that 30\% of the galaxies are probably outside the redshift limits of their respective redshift bins. If instead we choose to cut galaxies at $3\sigma_{68\%}$ from the redshift bin limits,  we only decrease this probability to 28\%.  (iii) It gives us theoretical predictions that better reproduce the measurements. Summing the pdfs produces redshift distributions with tails falling outside the redshift bins limits. As a result, we predict projected power spectra with larger amplitudes at small and large scales. They reproduce better the measurements.

Second, we must emphasize that the redshift distributions we obtain are weighting functions that we use to project the 3D matter power spectrum along the line of sight, and predict the behavior of several estimators. In particular, we use the lensing efficiency in place of the lensing-effective matter distribution. This latter would be the true matter redshift distribution multiplied by the lensing efficiency. Nonetheless, by averaging over many line of sights, we expect the lensing efficiency to approximate the lensing-effective matter distribution, and thus produce accurate predictions.

Third, the centers of foreground bins F4 and F5 do not perfectly match the peak of the lensing efficiency for catalog B3. In section~\ref{sec:discussion}, we show how
such a mismatch can alter the bias measurements. We show that the effects are particularly severe at small scales in the non-linear regime. Fortunately for bins F4 and F5 most of the angular scales we probe correspond to physical scales in the quasi-linear and linear regime. Since the
scale-dependence of the bias in the linear regime is small, the
effect of mismatch in redshift will be small. 

\subsubsection{Catalog cuts by flux and stellar mass}

In order to investigate the dependence of bias with stellar mass, we further partition our foreground catalogs into one flux-selected and two stellar-mass selected catalogs. 
The COSMOS catalog for the foreground galaxies is complete up to
$I_{814W} < 26.5$ and $K_s < 24$ in AB magnitude \citep{capak2007,ilbert2009,leauthaud2007}.
The limit in $K_s$-band is more conservative and comes from estimating the observed magnitude of a maximal $M_*/L$ stellar population model with solar metallicity, no dust, and a $\tau=0.5\ \mathrm{Gyr}$\ burst of star formation at redshift $z_\mathrm{form}=5$ \citep{leauthaud2011b}. 
\citet{bundy2010} computed the 80\% completeness
in stellar mass for magnitude selected galaxies $K_s < 24$ and
$I_{814W} < 26.5$. He found the 80\% completeness lower limits in log $M_* /
\Msol$ to be 8.8, 9.3, 9.7 and  10.0 for his redshift bins [0.2,0.4],
[0.4,0.6], [0.6,0.8] and [0.8,1.0].  In order to have enough galaxy per bin, we bin in the two stellar mass bins  $10^9\ < M_*
< 10^{10}\ h^{-2}_{72}\ \Msol$ and $10^{10} < M_* < 10^{11}\
h^{-2}_{72}\ \Msol$. Our low stellar-mass galaxy samples is therefore
more than 80\% complete in all redshift bins, and our high stellar-mass galaxy sample is more than 80\%
complete in bins F1, F2, about 80\% complete in bins F3, and less than
80\% complete in bins F4 and F5. For comparison, the mean stellar mass
in our flux-limited galaxy samples in bins F1, F2, F3, F4 and F5 are
9.6, 9.7, 9.8, 10.0 and 10.1 in units of $\log(M_*/h^{-2}_{72}\
\Msol)$.

\begin{table}
    \centering
    \small
    \caption{Redshift range and number of galaxies per foreground and background bins.}
    \begin{tabular}{lcccc}
        Bin & Redshift & \multicolumn{2}{c}{Stellar Mass Range [$h_{72}^{-2}\ \Msol$]} & \\
                & Range & $10^9 \to10^{10}$ & $10^{10} \to 10^{11}$ & All \\
        \tableline
        \tableline
        F1 &  [0.17, 0.27] & 1,298  & 644 & 7,920 \\
        F2 &  [0.32, 0.42] & 3,151  & 1,770 & 13,898 \\
        F3 &  [0.47, 0.57] & 3,215  & 1,456 & 11,256 \\
        F4 &  [0.60, 0.75] & 7,776  & 3,899 & 22,958 \\
        \vspace{5pt}
        F5 &  [0.87, 1.07] & 11,828 & 6,223 & 27,911 \\
        B1 &  [0.3, 0.8]   & ---  & --- & 63,529 \\
        B2 &  [0.8, 1.4]   & ---  & --- & 70,157 \\
        B3 &  [1.4, 4.0]   & ---  & --- & 59,056 \\
        \tableline
    \end{tabular}

    \label{tab:bins}
\end{table}

\section{Lensing Method}
\label{sec:formalism}

In gravitational lensing theory, the mass of foreground structures locally perturbs space-time, deviates the path of light-rays coming to us from background galaxies, and distorts their intrinsic shape. By analyzing the shape of many background galaxies, it is therefore possible to infer the properties of the intervening structures. For instance, the correlation of ellipticity measurements of background galaxies tells us about the clustering of the foreground structures. As such, gravitational lensing is a direct probe of the matter density field
projected along the line of sight. Perturbations of the background galaxy shapes are of two kinds. The convergence $\kappa$ quantifies the amount of isotropic amplification, and the shear, $\gamma = \gamma_1 + i \gamma_2$ quantifies the amount of stretch along some direction. 

If we consider a sample of background galaxies with lensing efficiency as a function of redshift $g(z)$, the convergence and shear produced by the matter along the line of sight, expressed in terms of   density contrast $\delta$ is

\begin{equation}
    \kappa(\btheta) = \int^{w_h}_0 \d w\, g(w) \delta(f_K(w)\btheta, w)\,,
\end{equation}
\begin{equation}
    \gamma(\btheta) = -\frac{1}{\pi} \int \d^2\btheta'\,
    \kappa(\btheta'-\btheta)\frac{1}{(\theta_1' - \theta_2')^2}\,.
\end{equation}

\citet{schneider1998a} introduce the mass aperture statistic $M_{ap}$ defined as

\begin{equation}
    \mathrm{M}_{ap}(\theta) = \int_0^\theta \d^2 \vartheta\,
    U(|\vartheta|)\kappa(\vartheta)\,,
\end{equation}

\noindent where the integral extends over a disk of angular radius
$\theta$. $U(\theta)$ is a compensated weight function
which vanishes for $\vartheta > \theta$. In a similar manner, the
aperture number count of galaxies at a scale $\theta$ is defined as

\begin{equation}
    N(\theta) = \int_0^\theta \d^2 \vartheta\, U(|\vartheta|) \delta_g(\vartheta)
\end{equation}

\noindent where $U(\vartheta)$ is the same function as above, and
$\delta_g(\theta)$ is the galaxy density contrast.

\subsection{Variance aperture statistics}

\citet{schneider1998b} and \citet{vanwaerbeke1998} developed a
formalism based on aperture statistics to measure the dependence of
bias with scale. They define the following aperture statistics from
the auto- and cross-power spectra of galaxies and matter $P_n(\ell)$,  $P_
\kappa(\ell)$ and $P_{n\kappa}(\ell)$ respectively

\begin{equation}
    \label{eq:nap}
    \nap = 2\pi \int_0^{\infty} \d \ell\ \ell\ P_n(\ell)
   \left[I(\ell\theta)\right]^2\,,
\end{equation}
\begin{equation}
    \label{eq:map}
    \map = 2\pi \int_0^{\infty} \d \ell\ \ell\ P_\kappa(\ell)
   \left[I(\ell \theta)\right]^2\,,
\end{equation}
\begin{equation}
    \label{eq:nmap}
    \nmap = 2\pi \int_0^{\infty} \d \ell\ \ell\ P_{n\kappa}(\ell)
    \left[I(\ell\theta)\right]^2\,.
\end{equation}

\noindent The filter $I(x)$ is defined as \citep{schneider1998a, hoekstra2002,simonp2007}

\begin{equation}
    I(x) = \frac{12}{\pi} \frac{J_4(x)}{x^2}
\end{equation}

\noindent where $J_4(x)$ is the Bessel function of order 4. This filter is a narrow band filter, which  peaks at $x\sim 4.25$. The power-spectra $P_n(\ell)$,
$P_\kappa(\ell)$ and $P_{\kappa n}(\ell)$ derive from the 3D matter power-spectrum $P_m(k,w)$ projected along the line of sight,  weighted by the functions $g(w)$,  $p_f(w)$ or
 $p_f(w)g(w)$, and altered by the bias parameters $b$ and $r$. Their respective analytical expressions are 

\begin{equation}
    \label{eq:pn}
    P_n(\ell) = \int_0^{w_h} \d w\, \frac{[p_f(w)]^2}{[f_K(w)]^2} b^2
    P_m \left( \frac{\ell}{f_K(w)}, w \right) \,.
\end{equation}

\begin{equation}
    \label{eq:pk}
    P_\kappa(\ell) = \int_0^{w_h} \d w\, \frac{[g(w)]^2}{[f_K(w)]^2}
    P_m \left( \frac{\ell}{f_K(w)}, w \right) \,,
\end{equation}

and

\begin{equation}
    \label{eq:kn}
    P_{\kappa n}(\ell) = \int_0^{w_h} \d w\, \frac{g(w)
    p_f(w)}{[f_K(w)]^2} b r P_m \left( \frac{\ell}{f_K(w)}, w \right) \,.
\end{equation}

We use the Limber approximation here  to simplify the integrals.

The technique developed in this paper is powerful because all the power spectra are filtered by the same narrow band filter $I(x)$. In contrast, the estimators $\xi_+$, $\xi_-$,  $\omega(\theta)$ and $\gamt$ involve four different filters (see Appendix~\ref{sec:correlators}). Combining them directly would lead to a mixing of modes in Fourier space, and would hamper the interpretation. 

\subsection{Bias measured from aperture statistics}
\label{sec:f1f2}

\begin{figure}
    \centering
    \includegraphics[width=\linewidth]{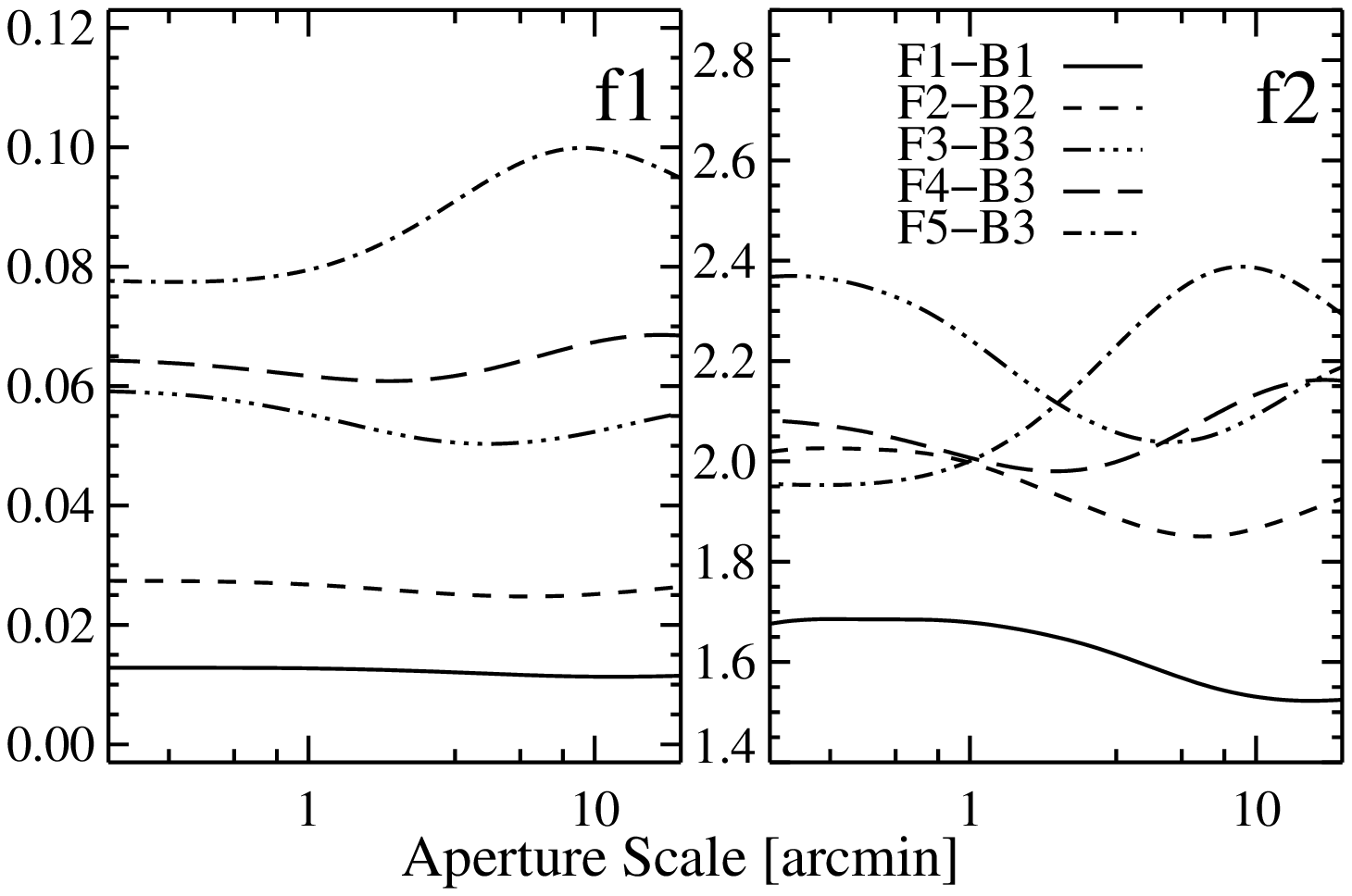}
    \caption{Bias calibration factors $f1$ and $f2$ used to calibrate $b$
    and $r$ respectively. We compute them assuming a \citet{smith2003}
    non-linear power-spectrum with unbiased foreground galaxies ($b=r=1$), a \citet{eisenstein1999} transfer function, and the WMAP7
    cosmological parameters \citep{komatsu2010}.
    Deviations from a straight line are due to mismatches in Fourier
    space of scales probed by the galaxy and the matter aperture
    statistics $\nap$ and $\map$ respectively. A straight line would
    indicate a perfect match between the two statistics. }

    \label{fig:f1f2}
\end{figure}

The bias parameters $b$ and $r$ are defined as  ratios of
the variance aperture statistics  

\begin{equation}
    \label{eq:bf1}
    b(\theta) = f_1(\theta, \Omega_m, \Omega_\Lambda) \times
    \sqrt{\frac{\nap}{\map}}\,,
\end{equation}

\begin{equation}
    \label{eq:rf2}
    r(\theta) = f_2(\theta, \Omega_m, \Omega_\Lambda) \times
    \frac{\nmap}{\sqrt{\nap \map}}\,.
\end{equation}

The functions $f_1$ and $f_2$  correct for the fact that different cosmological volumes are probed by the different statistics. They are defined as

\begin{equation}
    f_1(\theta) = \left. \sqrt{\frac{\map}{\nap}} \right|_{r=b=1}\,,
\end{equation}

\begin{equation}
    f_2(\theta) = \left. \frac{\sqrt{\nap\map}}{\nmap} \right|_{r=b=1}\,,
\end{equation}

 \noindent and are computed assuming a \citet{smith2003}
    non-linear power-spectrum with unbiased foreground galaxies ($b=r=1$),  a \citet{eisenstein1999} transfer function,  and the WMAP7
    cosmological parameters \citep{komatsu2010}.

 In Figure~\ref{fig:f1f2}, we show the
evolution of $f_1$ and $f_2$ as a function of scale and redshift.
 These functions are not constant in scale. The amplitude of the deviations is larger for the higher redshift samples of galaxies. In Figure~\ref{fig:f1f2}, function $f_1$ for pair F1-B1 is nearly constant, whereas it varies by 25\% for pair F5-B3. The scale-dependence of $f_1$ and $f_2$ is due to the fact that many different physical scales are projected into the same angular bins, and the weighting functions in $\nap$, $\map$ and $\nmap$ are different. 

\citet{vanwaerbeke1998} showed that $f_1$ and $f_2$ do not depend
on the shape of the power-spectrum. We performed a set of tests to verify this statement, and reached the same conclusion.   We computed $f_1$ and $f_2$ with a power spectrum for the linear regime \citet{peacock1996} instead of  a \citet{smith2003} power-spectrum. We altered the amplitude $\sigma_8$ and spectral index $n_s$ by 20\%, and we measured less than a 5\% difference. The amplitude of the power spectrum does not have a strong influence on $f_1$ and $f_2$. 
 Note however that the calibration functions $f_1$
and $f_2$ scale with $\Omega_m$ \citep{simonp2007}. We 
fixed it to the WMAP7 value.

\subsection{Physical scales}

Because we make measurements in bins of angular separations rather than physical separation, the bias parameters we measure are averages of the true bias parameters  $b(k,w)$ and $r(k,w)$ in k-space and comoving distances. Their respective expressions are given by \citep{hoekstra2002}

\newcommand{\bavg}{\langle b \rangle}
\newcommand{\ravg}{\langle r \rangle}

\begin{equation}
    \label{eq:bavg}
     \bavg^2 (\theta) = \frac{\int_0^{w_h} \d w\ h_1(\theta,w)
    b^2(\frac{4.25}{\theta f_K(w)}, w)}{\int_0^{w_h} \d w\ h_1(\theta,w)}\,,
\end{equation}

\begin{equation}
    \label{eq:ravg}
    \ravg (\theta) = \frac{\int_0^{w_h} \d w\ h_3(\theta,w)
    r(\frac{4.25}{\theta f_K(w)}, w)}{\int_0^{w_h} \d w\ h_3(\theta,w)}\,.
\end{equation}

\noindent where the weighting functions $h_1(\theta,w)$ and $h_3(\theta,w)$ are respectively

\begin{equation}
    h_1(\theta,w) = \left( \frac{p_f(w)}{f_K(w)} \right)^2
    P_\mathrm{filter} (\theta, w)\,,
\end{equation}
\begin{equation}
    h_3(\theta,w) = \frac{p_f(w) g(w)}{[f_K(w)]^2} P_\mathrm{filter}
    (\theta, w)\,,
\end{equation}

\noindent and the filtered matter power spectrum is

\begin{equation}
    \label{eq:pfilter}
    P_\mathrm{filter}(\theta, w) = 2\pi \int_0^\infty \d\ell\ \ell\
    P_m \left(\frac{\ell}{f_K(w)}, w \right) \left[
    I(\ell \theta) \right]\,.
\end{equation}

Note that in Eq.~\ref{eq:bavg} and \ref{eq:ravg} the bias parameters $b(k,w)$ and $r(k,w)$ are estimated at scale $k=4.25 / \theta f_K(w)$. In these equations, it was assumed that the bias parameters are evolving slowly with scale, and their product by the $I(x)$ filter was equivalent to the product by a Dirac $\delta_D(x)$ function. Note as well that the functions $p_f(z)$ and $p_f(z)g(z)$ are narrow in redshift space. As a result, very few angular scales are mixed together when bias parameters $b(k,w)$ and $r(k,w)$ are projected along the line of sight. Therefore, the shapes of the bias parameters $\b$ and $\r$ in real space and $b(k,w)$ and $r(k,w)$ in k-space are very similar.

We use the weighting functions $h_1$ and $h_3$ to relate the physical scales to the angular scales at which bias parameters are measured. The scales in physical units are obtained via the following expressions

\begin{equation}
	\langle R_b(\theta)\rangle = 2 \pi \frac{\int_0^{w_h} \d w\, h_1(\theta,w)}{\int_0^{w_h} \d w\, h_1(\theta,w) \frac{4.25}{\theta f_K(w)}}
\end{equation}

\begin{equation}
	\langle R_r(\theta)\rangle = 2 \pi \frac{\int_0^{w_h} \d w\, h_3(\theta,w)}{\int_0^{w_h} \d w\, h_3(\theta,w) \frac{4.25}{\theta f_K(w)}}
\end{equation}
	
\begin{figure}
    \centering
    \includegraphics[width=\linewidth]{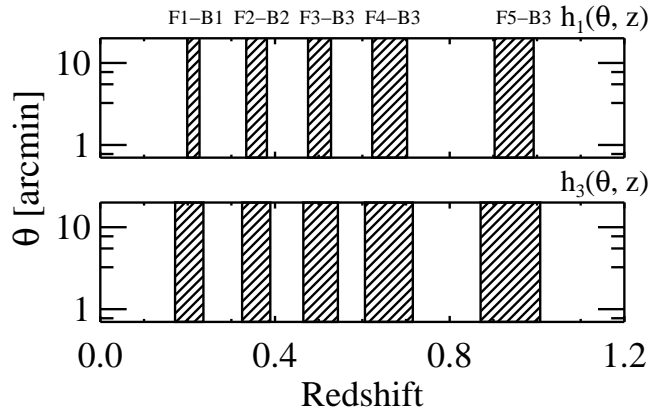}
    \caption{The 5 weighting functions $h_1(\theta,z)$ and $h_3(\theta,z)$ used to compute the analytical predictions of the bias parameters $\b$ and $\r$ for each bin Fi-Bj.  In both panels, the 5 hatched areas
    contain 68\% of the integrated signal in $h_1$ and $h_3$. The width of the hatched regions is scale-independent. }

    \label{fig:h1h3}
\end{figure}

 In Figure~\ref{fig:h1h3}, we show the weighting functions $h_1$ and $h_3$ for our 5 different pairs of redshift bins. The bias we measure in the following is the averaged bias of the galaxies in the hatched regions. The $h_1$ and $h_3$ functions extends over the whole redshift range, but we only hatched the regions containing 68\% of integrated signal in $h_1$ and $h_3$. We found the  widths of the hatched regions to be scale independent.  

\section{Simulations}
\label{sec:simulations}

\begin{figure}
    \centering
    \includegraphics[width=\linewidth]{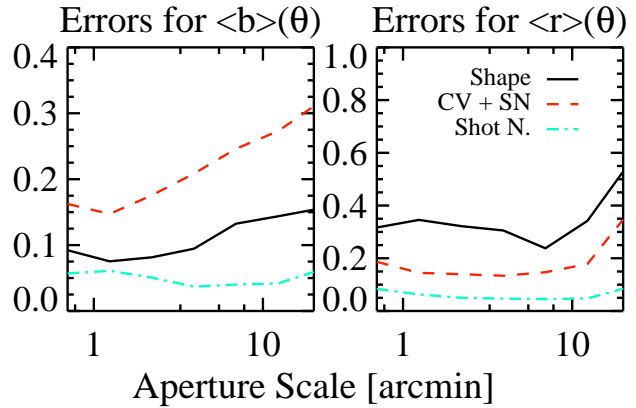}
    \caption{The error budgets for the galaxy bias parameters $\b(\theta)$ and
    $\r(\theta)$, obtained for foreground F1 and background B1. Shot
    noise and cosmic variance are obtained from simulations,
    whereas shape noise comes from the real data. }

    \label{fig:errors}
\end{figure}

We use N-body simulations to quantify the effect of cosmic variance on our bias measurements. 
Theoretical estimates of  the cosmic variance for mass aperture statistics exist
\citep{schneider2002a,joachimi2008}, but they break down in the
non-linear regime, making simulations a necessity. These simulations will also help us to check our tools. Indeed, the mock galaxies in these simulations are pure dark matter particles, and as such  perfect tracers of the underlying dark matter. With our tools, we should therefore find $b=r=1$ at all scales, and all redshifts. Any deviation would highlight a flaw in our tools, or an artifact in the method.

We generate a set of 7 light-cones of 10x10 square degrees each, and
extending from redshift $z=0$ to $z=4$. The particle mass is
$7.5\times10^{10}\ \Msol$. We estimate the shear and convergence on a
3D grid (fine in the angular direction, coarse in the radial
direction), using a no-radial-binning method \citep{kiessling2010}.

We make sure that the redshift distribution of the particles matches a
\citet{efstathiou1991} redshift distribution $p(z) \propto z^2 \exp[-(
z / z_0)^2 ]$, with $z_0 = 0.78$ and 66
galaxies per square arc-minute.
 This redshift distribution is close to
the actual redshift distribution in COSMOS \citep{massey2007b}. Finally,
each simulated catalog is cut into 36 COSMOS-size catalogs, and we
remove masked galaxies in the same way as we do with the real data. We obtain 343 mock catalogs of about 400,000 galaxies
each.  We bin these galaxies in redshift, and randomly resample them so that the number of  foreground and background galaxies match the numbers in Table~\ref{tab:bins}. The redshift of the  galaxies is computed analytically from their comoving distance assuming a WMAP7 cosmology. To keep it simple, we do not simulate photometric redshift uncertainties.

Figure~\ref{fig:errors} shows the error budget for the bias $b$ and the
correlation coefficient $r$ computed for the foreground redshift bin F1,
and the background bin B1. We split the error budget into cosmic
variance, shot noise and shape noise.
 To quantify the uncertainty in our results due to  cosmic variance, we compute the bias parameters for each COSMOS realization. The variance of the resulting distribution is therefore due to cosmic variance and shot noise. To quantify the amount of shot noise only, we bootstrap the galaxies in one single COSMOS realization. 

As shown in Figure~\ref{fig:errors}, we find the cosmic variance for parameter
$\b$ computed in foreground bin F1 and background bin B1 to be larger
than the shape noise. In contrast for the correlation coefficient $\r$, shape noise dominates at all scales. $\r$ is more sensitive to shape
noise because it contains the contributions from $\nmap$ and $\map$,
which are both affected by shape noise.  We perform the same analysis with the other pairs in bins Fi-Bj, and find that shape noise is always larger than cosmic variance both for bias parameters $\b$ and $\r$. The reason is that background galaxies in bins B2 and B3 are fainter, so their shape estimation is more noisy. 

We also used the simulations to test our treatment of the photometric redshift, and especially the way we reconstruct the weighting functions $p_b(z)$ and $p_f(z)$ for the background and foreground redshift bins. We simulated photometric redshifts with different redshift uncertainties from $\d_z / (1+z) = 0.01$ to $\d_z / (1+z) =0.05$. We reconstructed the weighting functions of bins F1 and B1. For bin F1 with $\d_z / (1+z) = 0.05$, we measured the rms between the recovered and the true $p(z)$ to be $\mathrm{rms} =4.49\pm0.32$  for the standard histogram technique,  and $\mathrm{rms}=4.10\pm0.33$ for the sum of pdfs technique. For bin B1 with errors $\d_z / (1+z) = 0.01$, we measured $\mathrm{rms} = 0.48\pm0.01$ and $\mathrm{rms}=0.65\pm0.01$ respectively. Therefore, our technique is better for large redshift uncertainties, and lower redshift bins. In our particular case, where redshift uncertainties are about $\d_z / (1+z) = 0.05$ for the faint part of the samples, our technique is globally as good as constructing a standard histogram of the best fit estimates. As already mentioned previously, we preferred this technique because galaxies outside the redshift limits were properly taken into account, which resulted in a better agreement between predicted and observed signals. To push this comparison further, more realistic galaxy redshift pdfs should be simulated,  but this is out of the scope of this paper.

Note finally that our estimates  of the systematic errors in our measurements of $\b$ and $\r$
might be underestimated because  we do not consider scatter due to inaccurate photometric redshifts, and we do not populate the dark matter
simulations with a realistic model of galaxies. Therefore, any
additional scatter introduced by physical processes  involved in galaxy formation will not be taken
into account. 

\section{Results}
\label{sec:results}

In this section, we present our measurements of the correlation
functions. We  use these measurements to derive the bias, and
 estimate its scale- and
redshift-dependence.

\subsection{Correlation functions}

As described in Appendix~\ref{sec:practical}, aperture statistics are
computed from auto- and cross-correlation functions. We use the
software
\textsc{Athena}\footnote{http://www2.iap.fr/users/kilbinge/athena},
which implements a tree-code, and computes auto- and cross-correlation
functions. For the tree-code, we choose an opening angle\footnote{In Athena, galaxy properties are  spatially averaged in nodes. The opening angle between a central and a tangential node is the ratio between the size of the tangential node divided by the distance between the nodes. The properties of two nodes are correlated if the opening angle is smaller than 0.04 rad.}  of 0.04 radians. All correlation functions are measured in 950 logarithmic
bins between 0.05 and 50 arc-minutes, which corresponds to a sampling scale of
$\Delta\vartheta = 0.007\vartheta$. The aperture statistics and the
bias parameters are also computed with this fine binning. We estimate the error bars with bootstrapping, i.e. we repeat the measurements 200 times with different galaxy samples randomly drawn from the input catalogs. For the figures, we rebin the data in coarse bins using a median binning technique. We used the median rather than the mean because the distributions of the bias parameters are not Gaussian, and so the median is more
robust.  However, this is a small correction since with the mean instead, the bias parameters $b$ and $r$ would be only about 3\% larger. To allow a straightforward comparison, we use the same binning for both the data and the simulations.

\begin{figure}
    \centering
    \includegraphics[width=\linewidth]{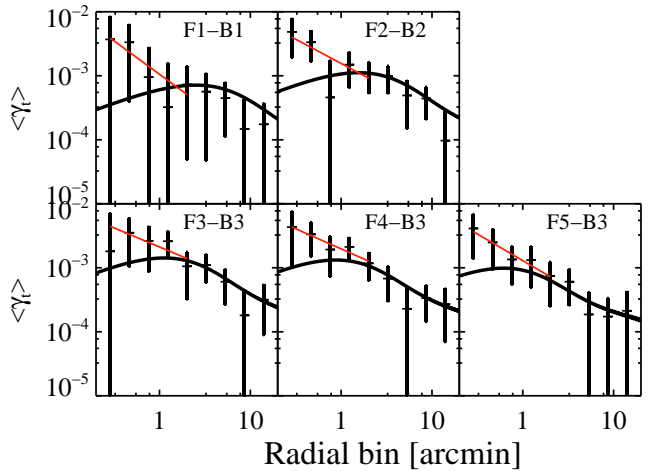}
    \caption{The mean tangential shear as a function of scale and redshift. The signal measured around random foreground galaxies has been subtracted to correct for systematic errors. To guide the eye, we overplot predictions
    derived from the \citet{smith2003} power-spectrum with unbiased galaxies (solid line).   We fitted power law functions to the data at small scales (red lines)  to assess the significance of the discrepancy with the predictions. Error bars include shape noise
  and shot-noise (thick line) and cosmic variance (thin line).}

    \label{fig:gamt}

\end{figure}

In the following, we will always compare our results to numerical predictions. We found the \citet{smith2003} fitting formula to give a good approximation of the measurements at large scales. In contrast at small scales, we found the fitting formula to under-predict the measurements (see e.g. Figures~\ref{fig:gamt}, \ref{fig:nap2} and \ref{fig:nmap}).  \citet{hilbert2009} had already noticed this discrepancy at  scales $\ell > 10000$, and suggested that it could be due to the low resolution of the simulations used by \citet{smith2003}. In our case, the discrepancies could at least partly also originate from the fact that we compare power-spectra  of biased galaxy samples and dark-matter. Indeed, power-spectra of galaxies are expected to deviate from pure dark matter power-spectra. Analyzing these deviations as a function of scale, redshift and galaxy properties is in fact the whole purpose of this paper. Therefore, comparing the measured and the predicted measurements already give us a first guess by eye of the amplitude of galaxy bias for a particular galaxy population.

\begin{figure}
    \centering
    \includegraphics[width=\linewidth]{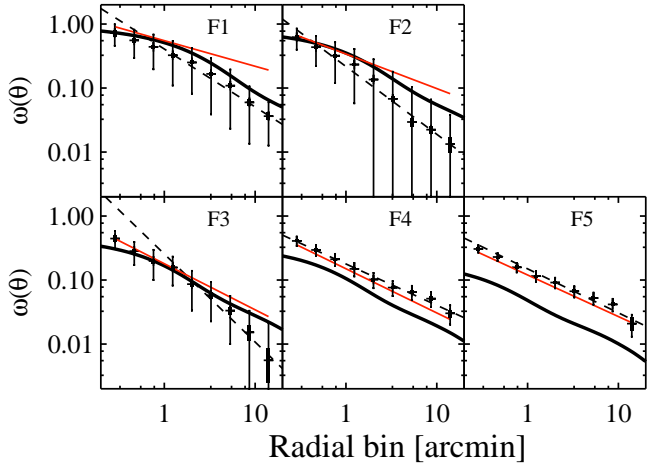}
    \caption{Projected galaxy auto-correlation for our flux-selected
    galaxy samples in bins F1 to F5.   (dashed line) Fit to the data with a power-law. The best fit scaling parameter $A_w$  and slope
    $\delta$ are reported in Table~\ref{tab:aw}   (red line) Power-law with parameters inferred from the fit to $\nap$. This represents the signal we should obtain after IC correction of $\omega(\theta)$ (see text). (bold line) Predicted signal derived from a \citet{smith2003}
    power-spectrum with unbiased galaxies. Error
    bars include shot noise (thick line) and cosmic
    variance (thin line). Cosmic-variance always dominates the measurements.   }

    \label{fig:w}
\end{figure}

\paragraph{Galaxy clustering} In Figure~\ref{fig:w} we show the projected
galaxy auto-correlation function $\omega(\theta)$ for our 5 foreground
redshift bins.  We compute $\omega(\theta)$ using the
\citet{landy1993} estimator defined as

\begin{equation}
    \omega(\theta) = \frac{DD}{RR} - 2\frac{DR}{RR} + 1\,.
\end{equation}

\noindent where the normalized numbers of pairs $DD$,  $RR$ and $DR$ refer to pairs of galaxy positions, random positions, and galaxy and random positions respectively.

By bootstrapping the data, we include shot noise in the error bars for the data, and cosmic variance for the simulations. At high redshift,
we calculate smaller error bars because (i) high redshift  bins contain more
galaxies, and (ii) we probe larger volumes and cosmic variance
decreases in larger volumes. In bins F4 and F5, we observe 2$\sigma$ and 3$\sigma$ deviations respectively between the measured and the predicted signal. These deviations already suggest that our flux-selected galaxy samples are biased with respect to the matter distribution (see Eq~\ref{eq:pn} and \ref{eq:w}). 

Galaxy auto-correlation functions have been found to closely follow a power-law \citep[e.g.][]{mccracken2007}. In order to assess the evolution of galaxy clustering with redshift,  we fit a power-law $\omega(\theta) = A_w
(\theta / 1')^{-\delta}$ to our measurements, and report the best fit parameters in
Table~\ref{tab:aw}. However, our estimator  $\omega(\theta)$ is mathematically known only up to an additive constant known as the integral constraint (IC), which depends on the volume and the mean galaxy density of the sample \citep{mccracken2007}. In the next section, we present how we correct for IC using the $\nap$ measurements.

\begin{table}
    \centering
    \caption{Scaling and slope of the power-law fit to
    $\omega(\theta)$ \sout{ and $\nap$}}

    \begin{tabular}{lcccc}
    \tableline
    \tableline
        & \multicolumn{2}{c}{Direct fit} & \multicolumn{2}{c}{IC Corrected using $\nap$}\\
        Bin & $A_w$(1') & $\delta$  & $A_w$(1') & $\delta$ \\
        \tableline
       F1 & 0.32 & 0.75 & 0.61$\pm$0.28 & 0.37$\pm$0.12 \\
       F2 & 0.21 & 0.93 & 0.50$\pm$0.16 & 0.41$\pm$0.09 \\
       F3 & 0.15 & 0.93 & 0.22$\pm$0.05 & 0.64$\pm$0.08 \\
       F4 & 0.18 & 0.65 & 0.21$\pm$0.03 & 0.54$\pm$0.05 \\
       F5 & 0.14 & 0.63 & 0.15$\pm$0.02 & 0.55$\pm$0.03 \\
        \tableline
    \end{tabular}
    \tablecomments{The IC corrected amplitudes $A_w$ and slopes $\delta$ are obtained from the
    fit to $\nap$, and Eq~\ref{eq:ic}.}
    \label{tab:aw}
\end{table}

\paragraph{Galaxy-Galaxy Lensing} In Figure~\ref{fig:gamt} we show the
 evolution with redshift of the  galaxy-galaxy lensing
measurements for our 5 flux-selected galaxy samples. 
  Overall, we observe our measurements to decrease with scale. To correct for systematic effects at large scales, we subtracted the signal measured around random foreground galaxies \citep{mandelbaum2005}. To guide the eye, we overplot the predicted signal derived from \citet{smith2003} power spectrum with unbiased galaxies. The match is  good at large scales, but we note a discrepancy at small scales. To assess the significance of this discrepancy, we compared the slopes of the data and the predictions at scales $\theta < 3'$. Taking into account the correlation between the data points, we fitted five power-law functions to the data at these scales, and found their slopes to be always negative at $2\sigma$ confidence level, hence at odd with the positive slopes of the predictions. This highlights  the limited accuracy of the \citet{smith2003} power-spectrum at small scales. 

\begin{figure}
    \centering
    \includegraphics[width=\linewidth]{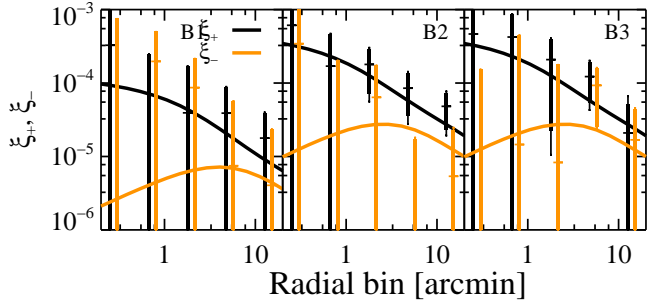}
    \caption{The shear-shear correlation function for the three background
    sources catalogs B1--B3.  Solid lines are the predicted signal with a
    \citet{smith2003} power-spectrum. Error bars include shape noise and shot noise (thick line),
    and cosmic variance (thin line). }

    \label{fig:xipm}
\end{figure}

\paragraph{Shear autocorrelation} Finally, we compute the shear
autocorrelation functions $\xi_+$ and $\xi_-$ for our 3 background
catalogs B1, B2 and B3.  They are presented in Figure~\ref{fig:xipm}.
The amplitude increases between B1 and B3 mainly because the lensing efficiency in Eq~\ref{eq:pk}  increases with redshift. The contribution of the structures along the line of sight is likely small, because they are mostly uncorrelated.


\subsection{Aperture statistics}

In this section we derive aperture statistics from the correlation functions. The
calculations are detailed in Appendix~\ref{sec:practical}.

\begin{figure}
    \centering
    \includegraphics[width=\linewidth]{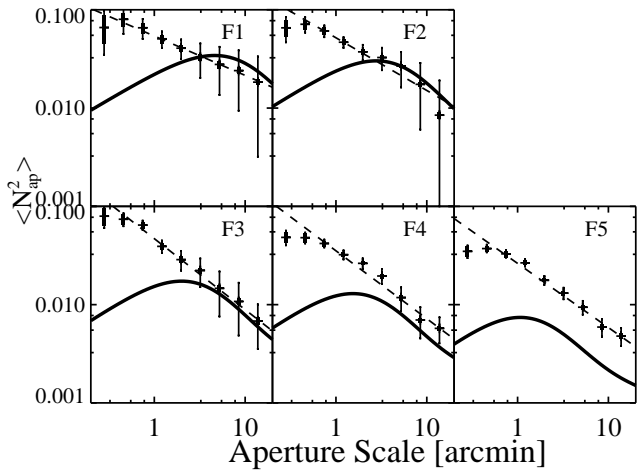}
    \caption{The galaxy aperture variance. To guide the eye, we overplot
    the predicted signal derived from a \citet{smith2003}
    power-spectrum with unbiased galaxies (solid line). As expected
    the prediction underestimates the signal at small scales. At
    large scales, the deviation in bins F4 and F5 already suggests that
    galaxies are biased (see also Figure~\ref{fig:b_all}).
    Error bars include shot-noise (thick line), and
    cosmic variance (thin line). }

    \label{fig:nap2}

\end{figure}

\paragraph{Galaxy aperture variance}   The measurements of $\nap$ for the 5 foreground bins
are shown in Figure~\ref{fig:nap2}. For each redshift bin, we fit a power-law. In contrast to $\omega(\theta)$, $\nap$ is not  affected by  IC, because the compensated filter $T_+$   which multiplies $\omega(\theta)$ in Eq~\ref{eq:napw}   cancels out any constant,  including  IC \citep{schneider1998a}.  \citet{simonp2007} derived an analytic relation between the parameters of a power-law fit to $\nap$ and the parameters of a PL fit to an IC corrected correlation function $\omega(\theta)$. The slope of $\omega(\theta)$ corrected for  IC is given by

\begin{equation}
   \label{eq:ic}
    f(\delta) = 0.0051 \delta^{11.55} + 0.2769 \delta^{3.95} + 0.2838
    \delta^{1.25}\,,
\end{equation}

\noindent where $\delta$ is the slope of the PL fit to  $\nap$. The amplitude of the PL fit to $\omega(\theta)$ is the same as the amplitude of the PL fit to $\nap$.
In Table~\ref{tab:aw}, we report theÊ IC-corrected amplitudes $A_w$ and slopes $\delta$ of the galaxy autocorrelation function $\omega(\theta)$  obtained with Eq~\ref{eq:ic}.

We find that the amplitude $A_w$ decreases with redshift, in agreement  with \citet{mccracken2007} who also found smaller amplitudes for fainter galaxies in COSMOS i.e. 
more likely located at higher redshift. Besides, we note a jump in amplitude between bins F3 and F2, i.e. between redshift $z\sim0.52$ and $z\sim0.35$. With respect to the slope $\delta$, we observe a slight but not significant (less than 1$\sigma$) steepening with redshift. Such a peculiar behavior could be explained by the large scale structure identified at redshift $z\sim0.7$ already detected in several other studies \citep{massey2007nat, guzzo2007, meneux2009, delatorre2010}. We guess that more massive galaxies in this structure would increase the slope because they are typically more clustered than average.

Finally, we observe an increasing deviation (more than $5\sigma$ in bin F5) between the measurements and the signal predicted with  a \citet{smith2003} power spectrum  and unbiased galaxies.   This suggests that galaxies are more biased at high redshift, in agreement with expectations. For bin F4 in the last 3 bins, we find on average a bias of about 1.4 between the data and the predictions, and for bin F5 in the last 5 bins, we find a bias of about 2.6.

\begin{figure}
    \centering
    \includegraphics[width=\linewidth]{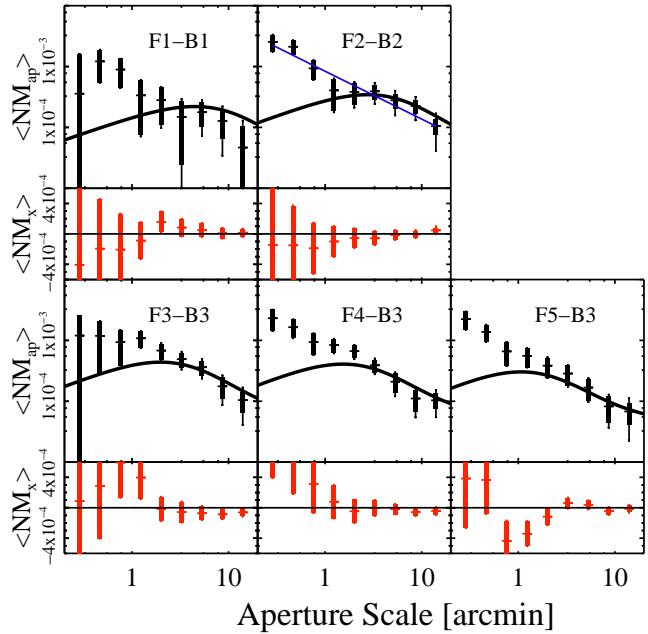}
    \caption{The galaxy-mass aperture covariance as a function of scale
    and redshift. To guide the eye at large scale, we also show the predictions derived from the  \citet{smith2003} power-spectrum with unbiased galaxies (solid line). For bin F2-B2, we found the change of slope of be not significant compared to a power-law fit (blue line).  In red,
    $\nmx$ quantifies the amount of ``B-modes'' (see text).  Error bars include
    shape noise (thick line) and cosmic variance (thin line).}

    \label{fig:nmap}
\end{figure}

\paragraph{Galaxy-mass aperture covariance}  The measurements of $\nmap$ for our 5
foreground redshift bins are shown in Figure~\ref{fig:nmap}. 
In agreement with the $<\gamma_t>$ measurements, we again note that a \citet{smith2003} power-spectrum under-predicts the data points at small scales. We performed a $\chi^2$-test, and found this disagreement to be significant in all bins at more than 3$\sigma$ (we included  the covariances matrices in the $\chi^2$-test). This disagreement is also present for the stellar-mass selected galaxy samples in Figures~\ref{fig:b_m9} and \ref{fig:b_m11}. 
At small scales in bin F2-B2, we  also note a slight change of slope at 1' scale. Again, we investigate the significance of this feature with a $\chi^2$-test, but found a reduced $\chi^2 = 1.60$ which is not enough to reject a simple power-law model. 

An E/B mode decomposition also exits for the $\nmap$ statistics. $\nmx$
quantifies the amount of B-modes in the measurement, and is computed by replacing $\gamma_t$ by $\gamma_x$ in Eq~\ref{eq:nmx}. $\gamma_x$ is the galaxy-galaxy clustering signal obtained with the coordinate system rotated by 45 degrees. This operation is commonly used to reveal B-modes contamination if a $\nmx$ signal is detected. 
Using a $\chi^2$-test and taking into account the correlation between the data points, we
find the B-modes for $\nmx$ to be consistent with zero in all bins Fi-Bj  at 95\% confidence level, hence confirming a very low level of contamination in our analysis.


\begin{figure}
    \centering
    \includegraphics[width=\linewidth]{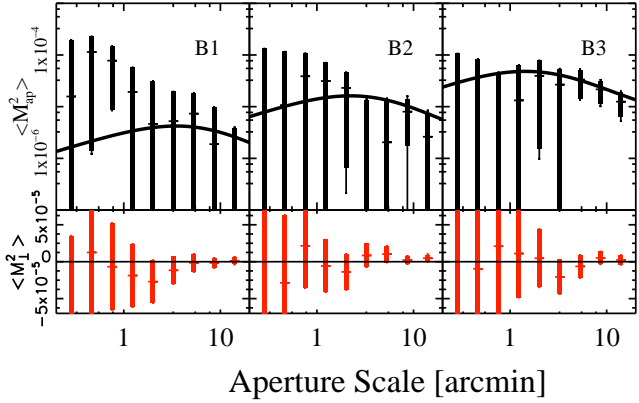}
    \caption{The mass-aperture variance as a function of scale and
    redshift.  To guide the eye, we overplot predictions derived from the 
    \citet{smith2003} power spectrum. In bin F2-B2, we fitted a power-law (blue line) to the data and found the change of slope of be not significant. (in red) The ``B-modes'' are consistent
    with zero at all scales in all bins. Errors include shape noise and shot noise (thick line), and cosmic variance (thin line). }

    \label{fig:map2}
\end{figure}

\paragraph{Mass aperture variance} The measurements of $\map$  for bins B1, B2 and B3 are shown in
Figure~\ref{fig:map2}. The solid line represents the predicted signal with a \citet{smith2003} power spectrum. We performed a $\chi^2$-test between the data points and the predicted signal and found that the data points are consistent with the predicted signal at 68\% confidence level for all bins B1, B2 and B3.  At all scales, shape noise
dominates over cosmic variance. Errors are smaller for
bin B3, because the signal is larger (about 10 times larger than the
signal in bin B1). 

\subsection{Bias of the flux-selected sample}
\label{sec:flux}

\begin{figure}
    \centering
    \includegraphics[width=\linewidth]{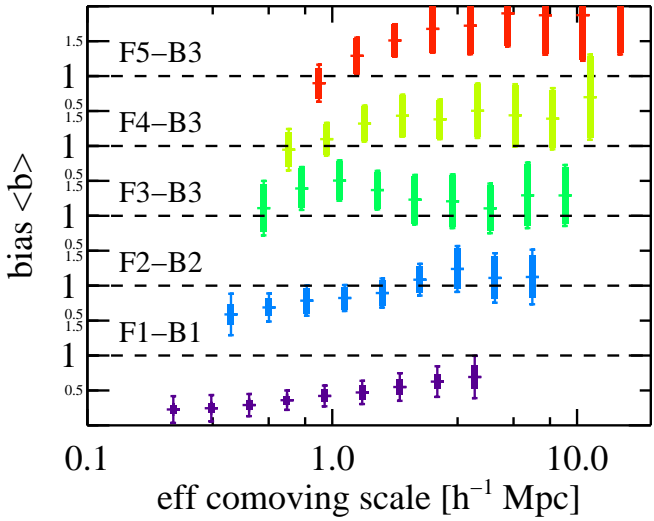}
    \caption{Evolution of bias in comoving scale and redshift for the
    flux-limited galaxy sample. Bias
    increases with redshift both because flux-selected galaxies reside
    in more massive halos, and because halos of any given mass are
    more biased at high redshift. At small scales, the change of slope can be interpreted as the transition between the two-halo and the one-halo terms in a halo model framework.  Errors bars include shape and shot noise (thick line), and cosmic variance (thin line). The horizontal dashed
    line marks the value $b=1$ for each bin. The data points are  correlated (see
    Section~\ref{sec:correlation}).  }

    \label{fig:b_all}
\end{figure}

\begin{figure}
    \centering
    \includegraphics[width=\linewidth]{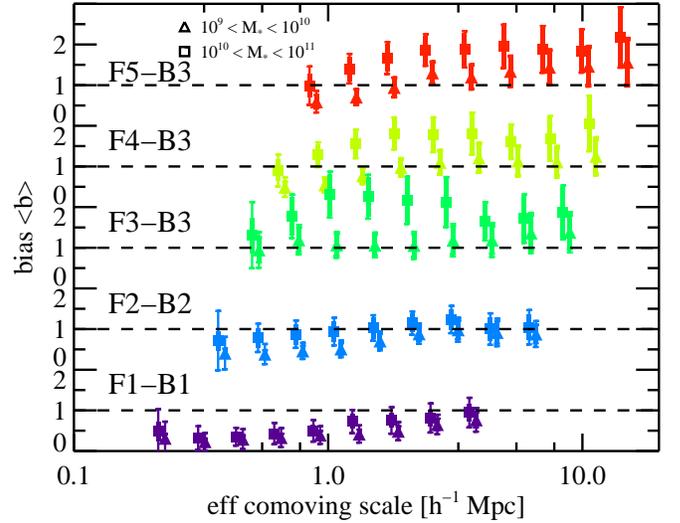}
    \caption{Same as Figure~\ref{fig:b_all}, but for stellar-mass selected galaxies. Bias
    increases with redshift and stellar-mass. The horizontal dashed
    line marks the value $b=1$ for each bin.}

    \label{fig:b_m}
\end{figure}

\subsubsection{Constant bias model}

In Figure~\ref{fig:b_all}, we show the galaxy bias $\b(R)$ for our
sample of flux-limited galaxies. We observe
that bias varies with scale and redshift, but since the data points are
correlated we  use a $\chi^2$-test to quantify the significance
of these variations. First, we assume the null hypothesis that
bias is scale and redshift-independent. To test this hypothesis, we
fit a constant $b_0$ to the 45 data points in the 5 redshift bins, and compute the $\chi^2$
statistics

\begin{equation}
    \label{eq:chi2}
\chi^2 = (\mathbf{B} - b_0)^T C^{-1} (\mathbf{B} - b_0)\,,
\end{equation}

\noindent where $\mathbf{B}$ is a vector containing the 45 data
points, and $C$ is their covariance matrix. To perform the fit and
find the best fit parameters, we use the IDL AMOEBA technique
\citep{nelder1965}, and repeat the process 100 times with different
starting values.  We obtain a best fit with $\chi^2=229$. Our data points are
correlated, so the effective number of degrees of freedom (dof) to
perform the $\chi^2$ test is less than the number of data points
$N=45$ minus the number of free parameters.  \citet{bretherton1999}
proposed several estimators for the number of degrees of freedom for
correlated data. We use the following
estimator\footnote{\citet{bretherton1999} mainly studied another
estimator $\mathrm{dof} = (\sum_i C_{ii})^2 / \sum_{ij} C_{ij}^2$,
which is equivalent if the data points are centered and normally
distributed with unit variance.}

\begin{equation}
    \label{eq:dof}
    \mathrm{dof} = \frac{N^2}{\sum_{ij} r^2_{ij}}
\end{equation}

\noindent where $r_{ij}$ is the correlation coefficient between data
points $b_i$ and $b_j$, and $N$ is the number of data points. Using
this estimator, we find $\mathrm{dof} = 32 - 1 = 31$, where we
subtracted 1 for the parameter we fit. The reduced $\chi^2$ is
therefore $\chi^2 / \mathrm{dof} = 7.4$, which allows us to
confidently reject a constant bias model given our data.

\subsubsection{Redshift-dependent model}

Here we discuss a test of the redshift dependence of bias. To our model, we
add the following redshift dependence 

\begin{equation}
    b(z) = b_0 + b_1 z\,.
\end{equation}

We obtain a reduced $\chi^2 / \mathrm{dof} = 2.5$. Although still not
a good fit, this model is nonetheless significantly better than the
previous redshift-independent model. 

So far, we have ruled out a constant bias with redshift and scale, but we did not isolate
the redshift dependence or scale dependence yet. This is the purpose of the rest of this
section.

\subsubsection{Scale-independent model}

Next, we  test the scale-dependence only. Our null hypothesis is
now that bias is scale-independent. For each redshift bin, we fit a
constant, and sum the $\chi^2$ for each individual redshift bin.  We
obtain a total $\chi^2 = 43$ for an effective number of degrees of freedom $\mathrm{dof} = 32 - 5 = 27$.  This means that there are 97.4\% chance that the scale-independent model is wrong.  Here we subtracted 5 for the 5 constant parameters we fit. A fit of this model to the stellar-mass
selected galaxy samples brings less stringent constraints, with $\chi^2 /
\mathrm{dof} = 1.49$ and $\chi^2 / \mathrm{dof} = 1.0$ for the low and high
stellar-mass selected samples respectively. 

\subsection{Bias of the stellar-mass selected sample}

In Figure~\ref{fig:b_m}, we show the galaxy bias $\b$ for  two
stellar mass-selected galaxy samples as a function of scale and
redshift. Previous studies have shown that stellar-mass is a good
tracer of halo mass \citep[see e.g.][and references
therein]{leauthaud2011b}, which in turn parametrizes most of the bias
models. In contrast, flux-selected samples are more affected by
selection effects. For instance, low surface brightness extended
galaxies are systematically under represented because they tend to
evade the magnitude cuts \citep{meneux2009}. Our stellar-mass selected samples should therefore tell us more about halo properties.

\subsubsection{Halo mass}

First, we  derive the average halo mass for our two samples of galaxies.  For this purpose, we
fit the bias model proposed by \citet{tinker2010}

\begin{equation}
    b(\nu) = 1 - A\frac{\nu^a}{\nu^a + \delta_c^a} + B\nu^b + C
    \nu^c\,,
\end{equation}

\noindent defined in terms of the peak height $\nu = \delta_c(z) /
D(z) \sigma(M)$, where $\delta_c(z)$ is the critical density for halo
collapse \citep{weinberg2003}, $D(z)$ is the growth factor, and
$\sigma^2(M)$ is the variance of the density fluctuations smoothed
with a top-hat filter of size $R = (3M / 4 \pi \bar{\rho}_m)^{1/3}$.
The parameters of this fitting function derive from fits to N-body
simulations. They are $A=1.0 + 0.24y \exp(-(4/y)^4)$, $a=0.44y -
0.88$, $B=0.183$, $b=1.5$, $C=0.019 + 0.107y + 0.19\exp(-(4/y)^4)$ and
$c = 2.4$. We define the parameter $y\equiv log_{10}\Delta$ for
overdensity $\Delta =200$ times the cosmological mean density.

Although this bias model is not explicitly a redshift-dependent model, we
make use of the fact that $\nu$ scales with redshift to test the redshift
dependence of our measurements.

The \citet{tinker2010} model is a halo bias model,
\ie\ it was calibrated on halo clustering measured in N-body simulations. In
contrast, we measure galaxy clustering for stellar-mass selected galaxies. Using it to infer halo mass
implies the two following assumptions~: (i) the dark-matter mass function is
constant in the range of halo mass we consider (ii) there is only one galaxy
per halo.  The first assumption is valid because our galaxies are embedded in
halos with peak height $\nu\sim 1$ and mass $\log_{10}(M/\Msol) \le 13$, and
the halo mass function in this regime is almost flat
\citep{sheth1999,tinker2008}.  The second assumption holds in the linear
regime where most of the signal comes from the clustering of central
galaxies.

To overcome the problem of having data points at different redshifts, we fit a
redshift-normalized peak height parameter $\nu_0$, defined such that $b(\nu_i)
= b(\nu_0 / D(z_i))$, where $z_i$ is the redshift of a data point $b_i$. Since
this model is only valid in the linear regime, we only consider the 16 data
points at scales $R>2\ h^{-1}\ \Mpc$. We find $\chi^2 = 3.13$ and $\chi^2 =
4.93$ for $\mathrm{dof} = 10.8$ and $\mathrm{dof}=11.2$ respectively for the
low and high stellar-mass galaxy samples. Therefore, the model proposed by \citet{tinker2010} fits well our data. 

This implies that our measurements agree with an increase of bias with redshift. In addition, we can derive the halo mass of our two stellar-mass selected galaxy samples. Indeed, the best fit values are $\nu_0 =
0.77^{+0.20}_{-0.31}$ and $\nu_0 = 1.01^{+0.24}_{-0.18}$ for the low and the high stellar-mass samples respectively. To compare to
\citet{leauthaud2011b}, we compute the peak height estimator at redshift $z=0.37$, $\nu_{z=0.37} = \nu_0 / D(0.37) = 0.93$ and
$\nu_{z=0.37} = 1.21$ for the low and high stellar-mass samples, respectively. This
translates into $9.64 <log_{10}(M_{200}/h^{-1}\Msol)<12.29$, and $11.51 <
log_{10}(M_{200}/h^{-1}\Msol) < 12.80$, respectively. These results are in very
good agreement with \citet{leauthaud2011b}, when considering that our two
stellar-mass selected samples range between $10^9\ h^{-2}\Msol < M_* < 10^{10}\
h^{-2}\Msol$ and $10^{10}\ h^{-1}\Msol < M_* < 10^{11}\ h^{-2}\Msol$.  As a last
check, we also fit the \citet{sheth2001} bias model, and found
similar results. The data are not stringent enough to distinguish the two
models.

\begin{figure}
    \centering
    \includegraphics[width=\linewidth]{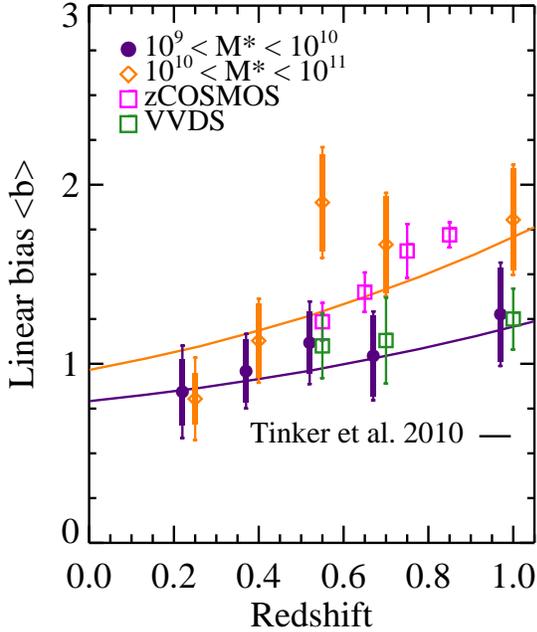}
    \caption{ Redshift evolution of bias averaged on scales $R>2\ h^{-1}\Mpc$ for the two stellar mass selected galaxy samples. Error bars include
shape noise, shot noise (thick line) and cosmic variance (thin line).  Solid lines are
the best fit predicted by the \citet{tinker2010} halo bias fitting function. At
$z=0.37$, the halo masses predicted by the model are $\log(M_{200}/h^{-1}\Msol)
= 11.58$ and $\log(M_{200}/h^{-1}\Msol) = 12.36$ for the low and high
stellar-mass selected galaxy samples.  Green boxes are measurements in VVDS for a
volume-limited ($\mathcal{M}_B < -20 + 5 \log h$) galaxy sample
\citep{marinoni2005}. Pink boxes are measurements in zCOSMOS for $M_B < -20 -
z$ galaxies \citep{kovac2009}. }

    \label{fig:bz_m}
\end{figure}

In Figure~\ref{fig:bz_m}, we show the evolution of bias with redshift. The data points are averages of points at scales $R > 2\ h^{-1}\Mpc$, and errors come from the respective total covariances matrices. The total covariance matrix is the sum of the data covariance matrix and the simulation covariance matrix. The shape noise, shot noise and cosmic variance are thus taken into account. For comparison, we
overplot the bias measured by \citet{kovac2009} (K09) with the zCOSMOS dataset,
and by \citet{marinoni2005} (M05) with the VLT VIMOS Deep Survey (VVDS).   The zCOSMOS dataset is the same as ours but with galaxies brighter than $I_{814} < 22.5$. The bias
measured by K09 follows our high stellar-mass galaxy samples, whereas the bias
measured by M05 follows our low mass samples. We attribute the difference in
bias between the two measurements to differences in galaxy selection functions.


We found bias in bin F3-B3 for
the high stellar-mass galaxy sample to be larger than expected. Although this data point still agrees with the bias model predictions  at 95\% confidence level, it shall be discussed further. Indeed 
at a similar redshift, K09 also identified a significant overdensity of galaxies which could explain the large bias value, but \citet{finoguenov2007} identified very few massive groups. These two inconsistent observations are puzzling. A halo model alike the one proposed by \citet{leauthaud2011a} could help us understand better the peculiar properties of the galaxies at this
redshift, in particular the fraction of satellite galaxies.

\subsubsection{Scale-dependent model}

In section~\ref{sec:flux}, we showed that a scale-dependent bias
model was preferred but still not significantly better than a
scale-independent model.  Nonetheless, we observe by eye that bias
systematically decreases at small scales, and the turn-down seems to
start at scale $R \sim 2\ h^{-1}\Mpc$, which could correspond to the
transition between the one-halo and the two-halo term, already
identified in several previous papers. According to \citet{zehavi2005},
the behavior of bias at this transition scale is due to pairs of
satellite galaxies. With a halo model applied to simulated data,
\citet{kravtsov2004} showed that a significant amount of pairs with
satellite galaxies was crucial for a smooth transition.
\citet{coupon2011}, using the CFHT Lensing Survey, and \citet{peng2011},
with SDSS data, also found that the fraction of satellite galaxies depend
on the \emph{overdensity} and not on the halo mass or redshift. Since
overdensities increase at late time, the number of satellite galaxies
increases in low redshift samples of galaxies. A change of slope in our
data would therefore indicate a change of the satellite fraction in the
galaxy samples.

In order to measure the significance of an evolution of the slope with redshift, we parametrize it as $\alpha(z)
= \alpha_1 + \alpha_2 z$. We also introduce a turn-down scale
$R_\mathrm{TD}$ beyond which bias evolves linearly. The final scale-dependent bias model we fit to the data is

\begin{equation}
    \label{eq:scalemodel}
    b(R,z) = b_\mathrm{lin}(\nu) \left\{ \begin{array}{lr}
        \left( \frac{R}{R_\mathrm{TD}} \right)^{\alpha_1 + \alpha_2 z} & R <
        R_\mathrm{TD}\\
        1 & R \ge R_\mathrm{TD} \end{array} \right. \,.
\end{equation}

\noindent where we use the redshift-dependent bias model
$b_\mathrm{lin}(\nu)$ of \citet{tinker2010} to describe the bias in
the linear regime.  We find $\chi^2 /\mathrm{dof} = 0.7$ and $\chi^2
/ \mathrm{dof}= 1.0$ for the low and high stellar-mass selected
samples respectively. The best fit parameters are reported in
Table~\ref{tab:scalemodel}. 

\begin{table}
    \centering
    \caption{Best fit parameters for the scale and redshift dependent
    model (Eq.~\ref{eq:scalemodel}) applied to the $10^9\ h^{-2}\Msol < M^* < 10^{10}\ h^{-2}\Msol$ (low) and  $10^{10}\ h^{-2}\Msol < M^* < 10^{11}\ h^{-2}\Msol$ (high) stellar-mass selected samples.}
    \begin{tabular}{lcc}
        \hline
        \hline
        & LOW & HIGH \\
        \hline
        $R_\mathrm{TD}\ h^{-1}\Mpc$  & $2.6\pm1.2$ & $1.0^{+0.8}_{-0.2}$ \\
        $\alpha_1$ & $0.42^{+0.32}_{-0.21}$ & $0.63\pm0.18$ \\
        $\alpha_2$ & $-0.17^{+0.44}_{-0.41}$ & $-0.44\pm0.24$  \\
        $\log{M_{200}/h^{-1}\Msol}$ & $11.7^{+0.6}_{-1.3}$ & $12.4^{+0.2}_{-2.9}$ \\
        $\chi^2 / \mathrm{dof}$ & 0.7 & 1.0 \\
        \hline
    \end{tabular}
    \label{tab:scalemodel}
\end{table}

The difference in 
 $\chi^2$ between the scale-independent (SI)
and scale-dependent (SD) model $\Delta \chi^2  = \chi^2_\mathrm{SI} /
\mathrm{dof}_\mathrm{SI} - \chi^2_\mathrm{SD} /\mathrm{dof}_\mathrm{SD}  = 0.79$ is not large
enough to completely rule out the SI bias model. 

The best fit parameters of the SD model merit some particular attention.
First, the scale at which bias turns down is detected to be between
$0.8\ h^{-1}\Mpc$ and $3.8\ h^{-1}\Mpc$ at 68\% CL. In contrast to what was observed in SDSS \citep{johnston2007}, the data
suggest $R_\mathrm{TD}$ to be marginally larger for less massive galaxies, but we
would need more bins in stellar-mass to confirm this tendency.

Second, the slope of bias at scales below $R_\mathrm{TD}$ is
 detected to be larger than zero, but at less than $2\sigma$ CL. We measured  $\alpha_1 = 0.42^{+0.32}_{-0.21}$ and
$\alpha_1=0.63\pm0.18$ for the low and high stellar-mass samples,
respectively.  Regarding its evolution with redshift, we found $\alpha_2 = -017{+0.44}_{-0.41}$ for the low stellar mass galaxy sample, hence no significant evolution, but $\alpha_2=-0.44\pm 0.24$ for the high stellar-mass selected sample, hence a slight flattening of the slope at small scales at higher redshift bins.

It is expected that the steepness of the slope is related to the occurrence
of satellite galaxies in our sample, with more satellite galaxies producing
shallower slopes. The coincident fact that the turn-down scale is larger
and the slope is shallower for the low stellar-mass sample suggests a
larger amount of satellite galaxies in our low stellar-mass selected
sample.  

\subsection{Correlation coefficients}

\begin{figure}[!t]
    \centering
    \includegraphics[width=\linewidth]{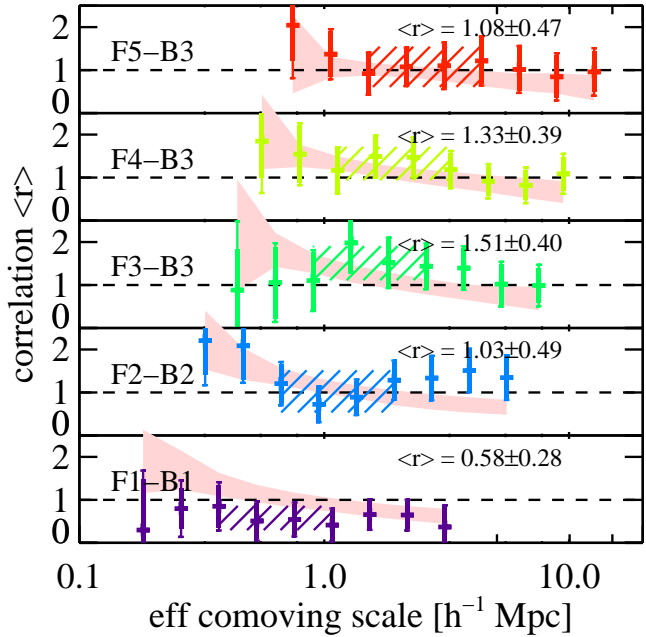}
    \caption{The correlation coeffcient $\r$ measured in redshift bins
    for our flux-limited galaxy samples. A deviation from $r=1$
    suggests the bias relation is stochastic, and not perfectly
    linear. Our method produces artificial deviations from $r=1$,
    which we quantify with pure dark matter simulations (pink
    area) for which we know a priori that $r=1$.  The size of the pink
    area is due to cosmic variance. The hatched area individuates the range of scales where the simulations are consistent with $r=1$. The averaged values of $r$ are computed in these areas.}

    \label{fig:r_all}
\end{figure}

In Figure~\ref{fig:r_all}, we show the correlation coefficient $\r$ for our
flux-limited galaxy samples. Our measurements agree with $r=1$ at all scales and all redshifts. We nonetheless observe some trend in the data, that might arise from possible artifacts in the method  (see Section~\ref{sec:discussion}). To highlight them, we applied the method to the simulated data, and indeed found a trend with a signal greater than 1 at small scale and lower than 1 at large scale. We show the correlation coefficient and  errors obtained from the simulations as pink regions. Consequently, the trends we observe in our data points can be explained by artifacts in the method. The pink hatched regions mark the scales where $r$ is consistent with 1 in the simulations. To compute the correlation factor for a particular galaxy sample, we only average the data points in these regions.  

We found no significant deviation from $r=1$, although in bin F1-B1, we obtain $r=0.58\pm0.28$.  \citet{hoekstra2001} and \citet{simonp2007}   found  the correlation coefficient $\r < 1$ at $8\sigma$, highlighting some stochasticity in the bias relations of their galaxy samples. In our case, we can only say that our flux-selected and stellar-mass selected samples are good tracers of
the underlying mass distribution given the error bars (see also Figures~\ref{fig:b_m9} and
\ref{fig:b_m11} for the results with the stellar-mass samples).

We have found no standard halo model to predict the scale-dependence of the correlation coefficient $r$. In contrast, \citet{neyrinck2005} propose a galaxy-halo model which does. In their model, they assume halos attach to galaxies, in contrast to the standard halo model where galaxies attach to halos. By construction, assuming all halos have the same density profile, the correlation-coefficient is always $r=1$. In the case halos have different density profiles, the correlation factor becomes scale dependent. At large scales, many halos are averaged over, resulting in a mean density profile. The matter distribution therefore matches the galaxy distribution, and $r=1$. At scales smaller than the minimum intergalactic distance,  $r$ drops below 1 because of the many different density profiles. Finally, assuming that the inner part of the density profiles are similar, $r$ raises back to 1.

In light of this model, we attempt to identify a dip in bins F1-B1 and F2-B2 at scale $R\sim 2\ h^{-1}\ \Mpc$. \citet{hoekstra2002} and \citet{simonp2007} also obtained such a dip at scale $R\sim 1\ h^{-1}\ \Mpc$ with much higher significance. We fit a constant $r=1$ to our data points in the hatched areas, and  measured $\chi^2 = 3.45$ in bin F1-B1 ($\mathrm{dof} = 2.75$, $>68$\% CL at $\chi^2>3.20$), and $\chi^2 = 2.20$ in bin F2-B2 ($\mathrm{dof} = 3.06$,  $>68$\% CL at $\chi^2 > 3.57$). A constant model $r=1$ therefore still provides a good fit  to the data.

\subsection{Bias correlation matrix}
\label{sec:correlation}

\begin{figure}
    \centering
    \begin{tabular}{cc}
    \includegraphics[width=0.5\linewidth]{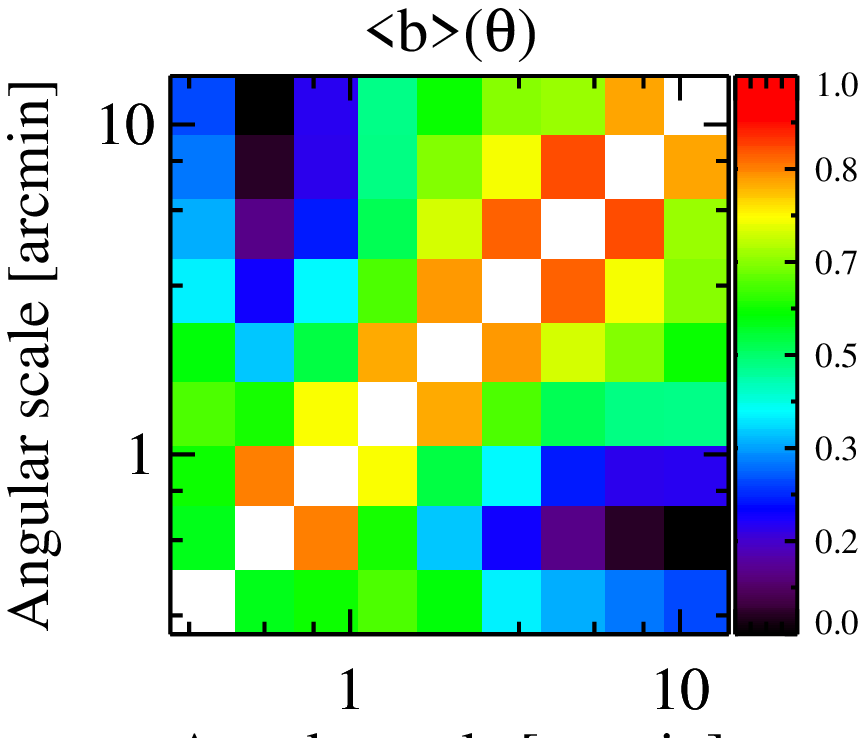}&
    \includegraphics[width=0.5\linewidth]{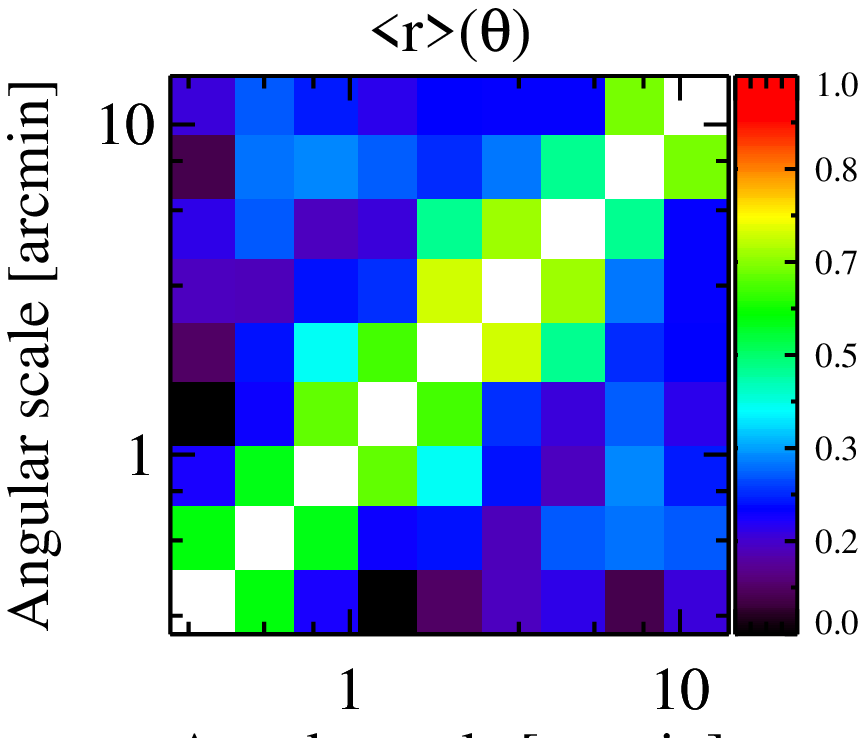}\\
    \end{tabular}

    \caption{ {\bf Left~:} The correlation matrix of estimator $\b(\theta)$
    for data points in Figure~\ref{fig:b_all} in bin F1-B1. {\bf Right~:}
    The correlation matrix of estimator $\r(\theta)$
    in Figure~\ref{fig:r_all} in bin F1-B1. Both correlation matrices include shape noise, shot noise and cosmic variance.  }

    \label{fig:b_cor} \end{figure}

Figure~\ref{fig:b_cor} shows the correlation matrix of the bias
(including cosmic variance) for the flux-limited galaxy sample F1-B1. The
data points are significantly correlated between 1' and 10' for
the bias parameter $\b$, and less correlated for the correlation
coefficient. The large amount of correlation at the smallest scales
$\theta < 0.4'$ is due to numerical artifacts (see section~\ref{sec:discussion}).

Note as well that the large correlation between the angular bins for
parameter $\b$ only shows up when the covariance matrix derived from simulations is added to the covariance matrix derived from
the data. Otherwise, the amount of correlation is weak because of the low signal-to-noise in the data.
The signal to
noise in the simulated data is much higher, hence the stronger signal
in the correlation matrix.

\section{Discussion}
\label{sec:discussion}

\subsection{Artifacts in the measurements}

It is well-known that aperture statistics, obtained through
integration of other estimators, are very sensitive to integration
limits. \citet{kilbinger2006} found that the signal is underestimated  by more than 10\% at
12 times the lower integration limit, i.e. 0.6' in our case. On the other
hand, at large scales, the signal is only valid up to half the upper scale
limit (see Eq.~\ref{eq:map}--\ref{eq:nap}). We used simulations
to assess the systematic effects produced by aperture
statistics. In Figure~\ref{fig:b_m9} and
\ref{fig:b_m11}, our simulations show that the correlation coefficient $\r$ is overestimated
by 10\% to 50\% at scales $\theta<1'$, depending on the
redshift bin. For the bias parameter $\b$,  deviations from $b=1$
occur at scales $\theta<0.7'$.

In the simulations, we also observed that the cross-correlation
signals $\nmap$  and $\gamt$ are biased low at large scales. Subtracting the signal measured around random foreground did not help in recovering $r=1$ at large scales. We found this
effect to be less important as the number of foreground galaxies
decreases with respect to the number of background galaxies. This can
be seen by comparing the red dashed lines in Figures~\ref{fig:b_m9} and
\ref{fig:b_m11}.  Although this effect is of minor consequences on our measurements given the size of the error bars, we tried to understand it. It is possible to demonstrate analytically that a constant additive systematic will cancel between sources 90 degrees separated. When foreground galaxies do not have sources all around them or the additive systematic is not constant, the average $\gamt$ is biased low. In the Sloan Digital Sky Survey, \citet{mandelbaum2005} found that subtracting the signal measured around random foreground galaxies was effectively removing this noise. However, our simulations do not include shape noise, and therefore the signal we measure at large scales cannot be due to shape noise.

Another source of systematic error could be related to the tree-code we use to compute the correlation functions. The documentation for the  tree-code
\textsc{Athena} mentions that choosing a too large opening angle can smear out the ellipticities of galaxies at large scales. Indeed, we find that if we increase the opening angle then the effect worsens, but it does not improve with values smaller than the one we use.  Because this systematic error is subdominant to our measurements, we decided not to further attempt its correction.


\subsection{Bin mismatch in redshift}

In this section, we  show that  a mismatch in redshift between peaks of the $p_f(z)$ and the lensing efficiency curves $g(z)$ can alter the bias measurements. We looked into this issue because this is the case in bins F4-B3 and F5-B3. To understand the origin of the problem, we should recall that in our method, the bias is the product of a function $f_1$ and a measurement $\nap / \map$. 

On the left panel of Figure~\ref{fig:mismatch}, we compute the ratios $\nap / \map$ and the inverse of the functions $f_1$ for different combinations of foreground and background sample bins. We find that between bins F1-B1 and F1-B3 $\nap / \map$ decreases by 37\% and 33\% at large and small scales respectively, whereas function $f_1$ decreases by 78\% and 64\%. As a result, our estimations of bias change. 

We do expect the signal to decrease because $\map$ increases with redshift, but we do not expect the measured and the predicted signal to change by different amounts. According to Eq~\ref{eq:bavg}, the bias measurements should only depend on the foreground redshift distribution, and not on the choice of the background galaxy sample and associated lensing efficiency. This is of course assuming that $\map$ and the matter power spectrum grows linearly, and the lensing-effective matter redshift distribution is well approximated by the lensing efficiency. For very wide surveys where many line of sights are averaged over, the lensing efficiency might be a good approximation of the effective matter distribution, but in COSMOS especially at low redshift, this might not be the case. On the other hand at small scales in the non-linear regime, the power spectrum might not grow linearly. In Figure~\ref{fig:mismatch},  smaller scales are more affected by bin mismatch.

Given the limited size of the COSMOS field, we therefore posit that cosmic variance or shape measurements could yield these discrepancies because the volumes probed by $\map$ are too small. 
This type of discrepancies warn us that for future surveys, we should try to match as closely as possible the mean redshift of the foreground galaxies and the peak of the lensing efficiency, in order to limit the effect of cosmic variance.  

\section{Conclusion}
\label{sec:conclusion}

The strength of the COSMOS survey is the exceptional quality of the
ACS imaging, and the 30-band photometry. Thanks to the latter, a
 photometric redshift can be derived for more than 600,000
galaxies at $I_{814W} < 26.0$ \citep{ilbert2009}. 

In this paper we made use of these two properties to investigate the
evolution of stochastic bias with scale and redshift in COSMOS. We partitioned our foreground galaxy catalogs in 3 categories~: one flux-selected catalog ($I_{814W} < 26.5$ and $K_s < 24$), and two stellar-mass selected catalogs. To estimate the bias parameter $b$ and the correlation coefficient $r$, we
applied a technique based on  aperture statistics described in
\citep{schneider1998b,hoekstra2002}. We used simulated
lensing catalogs derived from N-body simulations to quantify the
amount of error due to cosmic variance, and to test the method against
numerical artifacts. We found that weak lensing shape noise dominates
the error budget, and that we are affected by some numerical artifacts
at small scales ($\theta < 0.6'$), and at large scales for the
correlation coefficient $r$.

 Then, we used a $\chi^2$-test to assess the significance of redshift evolution and scale-dependence of bias. We found a significant  evolution of bias with redshift, but could not confidently rule out a scale-independent model. We found that both bias models proposed by \citet{tinker2010} and \citet{sheth2001} provide a good fit to the redshift evolution of our measurements at scales $R > 2\ h^{-1}\Mpc$. We used the result of the fit to derive the halo mass of our stellar-mass selected galaxy samples.  Finally, we observed that bias starts to decline below a scale of about $2\ h^{-1}\Mpc$. We proposed a bias model to describe this scale-dependence and obtained a very good fit with $\chi^2 / \mathrm{dof} \le 1$ for our two stellar-mass selected galaxy samples. We measured a  $3\sigma$ significance of the turn-down scale $R_\mathrm{TD}$ for the high stellar mass  sample, but could not draw conclusions about the evolution of the slope with redshift because of low signal to noise. 

 Finally, we observed that bias in bin F3-B3 is $2\sigma$ off the expectations with a \citet{tinker2010} bias model. We attribute this to the presence of large over-densities in the COSMOS field already
identified in \citet{kovac2009} and \citet{massey2007nat}. With respect to bias stochasticity parameter $\ravg$, our measurements are all consistent with  $\ravg = 1$. The linear bias model is  therefore a good fit to our data, albeit with somewhat large errors.

The COSMOS field is still too small to
produce decisive conclusions
about the evolution of bias, especially to characterize its
scale-dependence, and  to measure its stochasticity
in stellar mass-selected galaxy samples. On average, we measured signal to noise ratios $b / \sigma_b \sim 3$ and $r / \sigma_r \sim 0.5$. The most important source of error is shape noise in the background galaxies. \citet{schneider1998a} have shown that the error on $\map$ decreases as $1/N_b$. In order to increase the signal to noise by a factor of 10, we would need a field 10 times larger.  

 With their
RCS and VIRMOS-DESCART survey of 50.5 deg$^2$, \citet{hoekstra2002}
obtained signal to noise ratios about 8 on both bias parameters
$b$ and $r$. However,  they did not perform simulations to
disentangle systematic and statistical errors as we did in this paper. They also did not have redshift or stellar-mass estimates for their galaxies. Future ground-based wide field surveys such as KIDS and VIKING, DES, Pan-STARSS, LSST and HyperSuprime-Cam on Subaru should provide good photometric redshifts and stellar-mass to perform a more conclusive analysis of the scale- and redshift-dependence of bias.

 The main challenge for weak-lensing tomography with the future surveys (this bias analysis being one application) is to measure the shape of
background galaxies up to redshift $z=4$. This is essential to measure
the redshift evolution of galaxy properties and bias up to redshift
$z\sim 1$.  In their current implementation, the space-based missions Euclid and WFIRST propose a near-infrared (NIR) and a visible channel (for Euclid), and very wide field surveys. Visible and NIR  imaging are critical to produce good photometric redshifts \citep{jouvel2011}. These missions will also produce galaxy densities of about 30 per arc-minutes square. Therefore, with several millions of galaxies, they should be able to measure with  good S/N the scale-dependence of bias from very small to very large scales. 


\acknowledgments

We would like to thank the referee for his insightful comments that led to a significant improvement of the paper. It is also a pleasure to thank Chris Hirata and Fabian Schmidt for useful
discussions and helpful suggestions. We also gratefully thank Alexie
Leauthaud for her help and advice, and for providing the COSMOS lensing
catalog. This work was done in part at JPL, operated by Caltech under a contract for NASA. EJ acknowledges support from the Jet Propulsion Laboratory, in
contract with the California Institute of Technology, and the NASA
Postdoctoral Program. RM is supported by an STFC Advanced Fellowship. \copyright 2012. All rights reserved.

This work uses observations obtained with the
Hubble Space Telescope. The HST COSMOS Treasury program was supported through NASA grant HST-GO-09822. 

\bibliographystyle{aa}
\bibliography{/Users/ejullo/Library/texmf/bibtex/bib/all}

\appendix
\section{Practical estimators}
\label{sec:practical}

In practice, aperture statistics estimators are obtained from two
point correlation functions \citep{schneider2002a}.  The correlators
used in this work are~: (i) the angular correlation of the foreground
galaxy positions, $\omega(\theta)$, (ii) the mean tangential shear about
foreground galaxies, $\gamt (\theta)$, (iii) the shear-shear correlation
functions $\xi_\pm(\theta)$ as determined from the ellipticities of
the background galaxies \citep{simonp2007}~:

The angular correlation of the foreground galaxy positions is computed
with the \citet{landy1993} estimator defined as

\begin{equation}
    \omega(\theta) = \frac{DD}{RR} - 2\frac{DR}{RR} + 1\,.
\end{equation}

The mean tangential shear around foreground galaxies is computed in
angular scales, and is function of the shear components $\gamma_1$ and
$\gamma_2$

\begin{eqnarray}
    \gamt(\theta) &=&  \left\langle w^{(i)} \gamma_t^{(i,j)}
     \Delta^{(i,j)}(\theta) \right\rangle\\
    \nonumber
     &=&
    - \left\langle w^{(i)} \left( \gamma_1^{(i)} \cos{2
    \phi^{(i,j)}} + \gamma_2^{(i)} \sin{2 \phi^{(i,j)}}\right)
    \Delta^{(i,j)} \right\rangle
\end{eqnarray}

\noindent where $\phi^{(i,j)} = \theta^{(i)} - \theta^{(j)}$ is the
polar angle between foreground galaxy $i$ and background galaxy $j$.
The angle brackets stand for weighted mean over indices $i$ and $j$
and weights $w^{(i)}$.  $\Delta^{(i,j)}(\theta)$ is a binning function

\begin{equation}
    \Delta^{(i,j)}(\theta) = \left\{ \begin{array}{rl}
        1 &\mathrm{for}\ \theta \le |\theta_i - \theta_j| < \theta +
        \delta \theta \\
        0 & \mathrm{otherwise}
    \end{array}\right.
\end{equation}

The shear-shear correlation functions $\xi_\pm(\theta)$ are obtained
from the correlation of the shear components $\gamma_1$ and
$\gamma_2$~:

\begin{equation}
    \xi_+ (\theta)= \left\langle w^{(i)} w^{(j)} \gamma_1^{(i)}
    \gamma_1^{(j)} \right\rangle + \left\langle
    w^{(i)} w^{(j)} \gamma_2^{(i)} \gamma_2^{(j)} \right\rangle
    \Delta^{(i,j)}(\theta)
\end{equation}

\begin{equation}
    \begin{array}{rcl}\xi_-(\theta) & = & \left( \left\langle w^{(i)} w^{(j)}
    \gamma_1^{(i)} \gamma_1^{(j)} \cos{4\phi^{(i,j)}} \right\rangle \right. \\
    &-& \left\langle w^{(i)} w^{(j)}
    \gamma_2^{(i)} \gamma_2^{(j)} \cos{4\phi^{(i,j)}} \right\rangle
    \\
    & +& \left\langle w^{(i)} w^{(j)}
    \gamma_1^{(i)} \gamma_2^{(j)} \sin{4\phi^{(i,j)}} \right\rangle \\
    & +& \left. \left\langle w^{(i)} w^{(j)}
    \gamma_2^{(i)} \gamma_1^{(j)} \sin{4\phi^{(i,j)}} \right\rangle
    \right) \Delta^{(i,j)}(\theta)
\end{array}
\end{equation}

 Aperture statistics estimators can be expressed in terms of these
correlators \citep{schneider2002a,hoekstra2002}. First, the E- and
B-mode of the aperture mass variance $\map$ are obtained from the
shear-shear correlation functions $\xi_\pm(\theta)$

\begin{equation}
    \map = \frac{1}{2} \int_0^{2\theta} \frac{\d \vartheta\
    \vartheta}{\theta^2} \left( \xi_+(\vartheta)
    T_+\left(\frac{\vartheta}{\theta}\right) + \xi_-(\vartheta)
    T_-\left(\frac{\vartheta}{\theta}\right) \right)\,,
\end{equation}

\begin{equation}
    \mpe = \frac{1}{2} \int_0^{2\theta} \frac{\d \vartheta\
    \vartheta}{\theta^2} \left( \xi_+(\vartheta)
    T_+\left(\frac{\vartheta}{\theta}\right) +
    \xi_-(\vartheta) T_- \left( \frac{\vartheta}{\theta} \right)
    \right)\,.
\end{equation}

\noindent where

\begin{eqnarray}
    \nonumber
    T_+(x) & = & \frac{6(2-15 x^2)}{5} \left[ 1 - \frac{2}{\pi}
    \arcsin(x/2)\right] \\ \nonumber
    & & + \frac{x\sqrt{4-x^2}}{100\pi} (120 +2320x^2 - 754^4 \\
    & & + 132x^6-9x^8)
    \,,
\end{eqnarray}

\begin{equation}
    T_-(x) = \frac{192}{35\pi} x^3 \left( 1 - \frac{x^2}{4}
    \right)^{7/2}\,.
\end{equation}

$T_\pm(x)$ vanish for $x>2$. The B-mode aperture mass $\mpe$
provides a quantitative estimate of the systematics, since
gravitational lensing only produces E-modes.

Second, the aperture count variance $\nap$ is related to $\omega(\theta)$

\begin{equation}
    \label{eq:napw}
    \nap = \int_0^{2\theta} \frac{\d \vartheta\ \vartheta}{\theta^2}
    \omega(\theta) T_+ \left( \frac{\vartheta}{\theta} \right)\,.
\end{equation}

Finally, the galaxy-mass cross correlation aperture $\nmap$ is
obtained from the mean tangential shear $\gamt(\theta)$

\begin{equation}
    \label{eq:nmx}
    \nmap = \int_0^{2\theta} \frac{\d \vartheta\ \vartheta}{\theta^2}
    \gamt(\theta) F\left(\frac{\vartheta}{\theta}\right)\,
\end{equation}

\noindent   where the function $F(x)$ has no analytic expression and
also vanishes for $x>2$

\begin{equation}
    F(x) = 576 \int_0^2 \frac{\d t}{t} J_2(xt) \left[ J_4(t)
    \right]^2\,.
\end{equation}

\section{Correlators}
\label{sec:correlators}

We have seen in Eq~\ref{eq:nap} -- \ref{eq:nmap} that the aperture statistics $\nap$, $\map$ and $\nmap$ derive from their respective power-spectra multiplied by the same filter function $I(\ell \theta) = J_4(x) / x^2$. This filter is convenient because it is very narrow in Fourier-space, and almost acts as a $\delta_D(x)$ function. When integrated over the power-spectra, it results in very limited correlation between the Fourier modes. The aperture statistics are close approximations of the power-spectra but in real space. There is almost a bijective relation between angular scales, and Fourier modes. As a result, it is possible to combine the aperture statistics to derive the bias parameters as it is done in Fourier space. 

In practice, aperture statistics derive from the correlators  $\omega(\theta)$, $\gamt(\theta)$ and $\xi_\pm(\theta)$. Correlators also derive from the power-spectra, but filtered by different filters, as shown below \citep{hoekstra2002}

\begin{eqnarray}
    \label{eq:w}
    \omega(\theta) & = & \langle \delta n(0) \delta n(\theta)
    \rangle \nonumber \\
    & = & \frac{1}{2\pi} \int_0^\infty \d \ell\ \ell\ P_n(\ell)
    J_0(\ell \theta)\,,
\end{eqnarray}

\begin{eqnarray}
    \label{eq:gamt}
    \gamt(\theta) & = & \langle \delta n(0) \gamma_t(\theta) \rangle
    \nonumber \\
    & = & \frac{1}{2\pi} \int_0^\infty \d \ell\ \ell\ P_{\kappa
    n}(\ell) J_2(\ell \theta)\,,
\end{eqnarray}

\begin{eqnarray}
    \xi_\pm(\theta) & = & \langle \gamma_t(0) \gamma_t(\theta) \rangle
    \pm \langle \gamma_\times(0) \gamma_\times(\theta) \rangle
    \nonumber \\
    & = & \frac{1}{2\pi} \int_0^\infty \d \ell\ \ell\ P_\kappa(\ell)
    J_{0,4}(\ell \theta)\,,
\end{eqnarray}

Therefore, taking their ratio to derive the bias parameters is unreliable. Note as well that some filters alike the Besel function  $J_0(x)$ act as low-pass filters. Many Fourier modes are mixed together into one angular scale. Consequently, the correlation between the different angular scales is large. 

\begin{figure*}
    \centering
    \begin{tabular}{c}
        \includegraphics[width=0.9\linewidth]{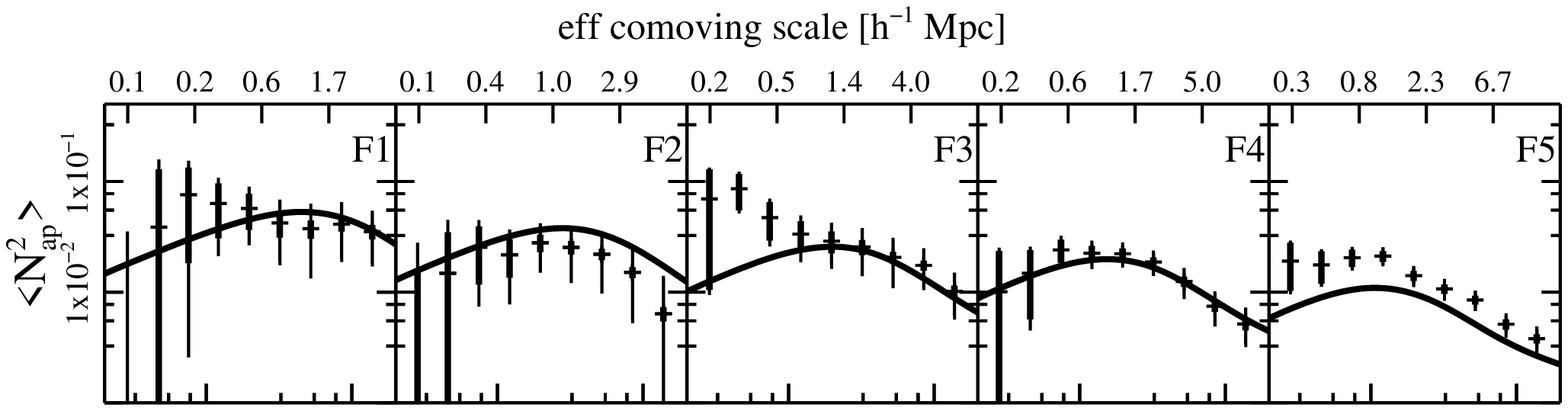}\\
        \includegraphics[width=0.9\linewidth]{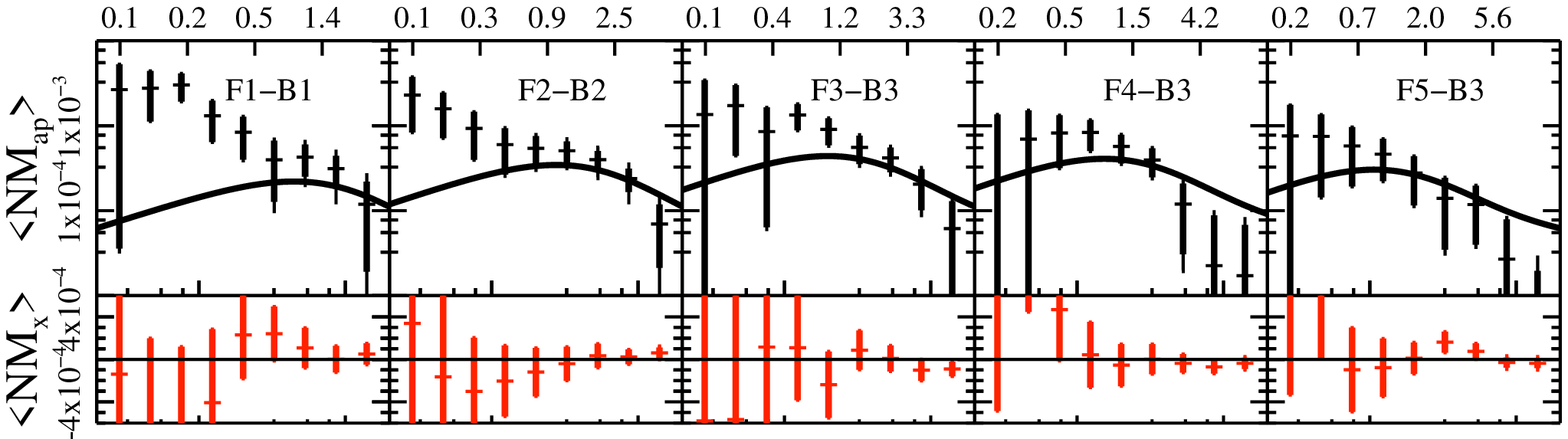}\\
        \includegraphics[width=0.9\linewidth]{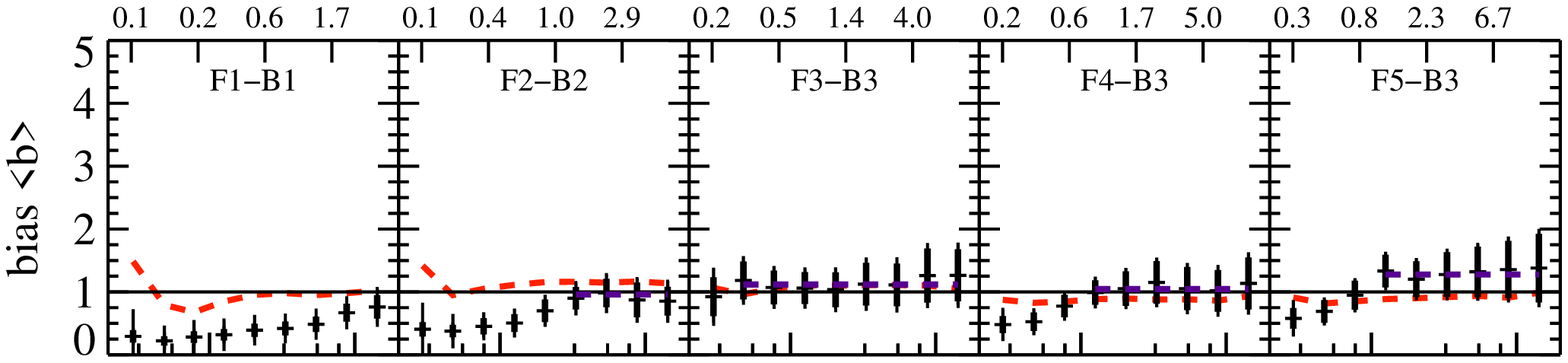}\\
        \includegraphics[width=0.9\linewidth]{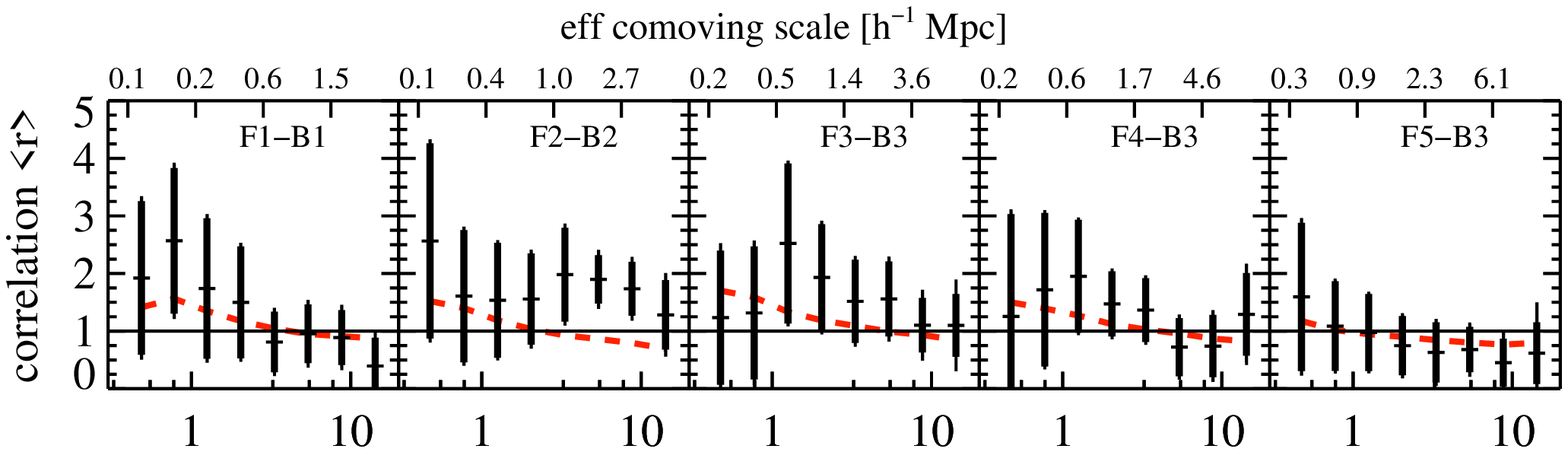}\\
    \end{tabular}
    \caption{Overview of the most important estimators involved in the
    measurement of bias and stochasticity for the stellar mass
    selected galaxy sample in the range $10^9 < M_* < 10^{10}\ h^{-2}\
    \Msol$. The black curve in the two upper panels is the theoretical
    prediction derived from a \citet{smith2003} power-spectrum, and it
    is meant to guide the eye at large scales. The discrepancies at
    small scales are expected (see text). The red dashed curves are the
    signals obtained in the simulations, which highlight numerical
    artifacts.  Data points must be interpreted with respect to these
    curves rather than $b=1$ and $r=1$.}

    \label{fig:b_m9}
\end{figure*}

\begin{figure*}
    \centering
    \begin{tabular}{c}
        \includegraphics[width=0.95\linewidth]{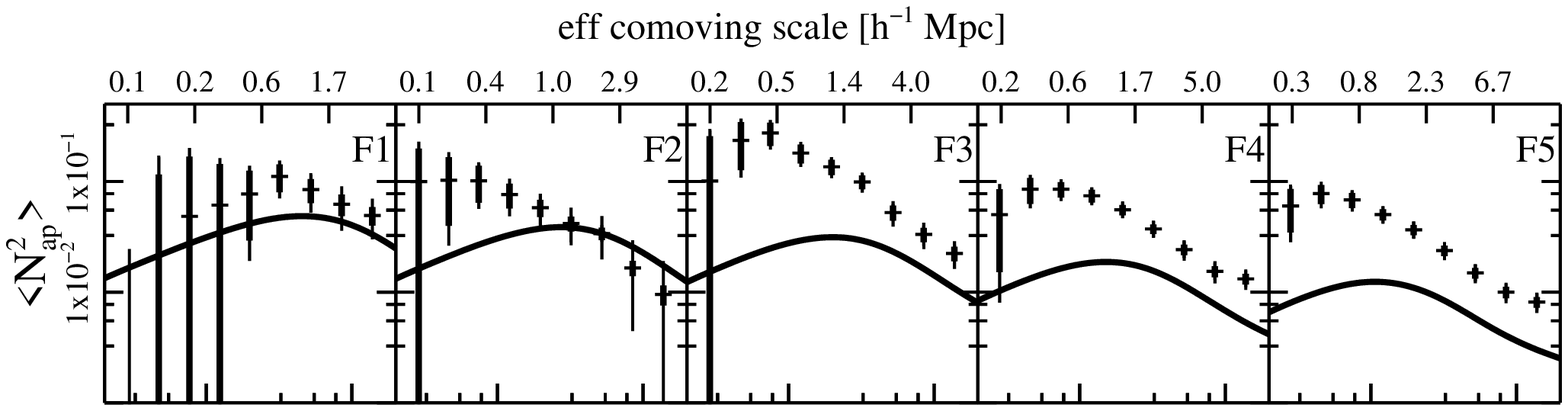}\\
        \includegraphics[width=0.95\linewidth]{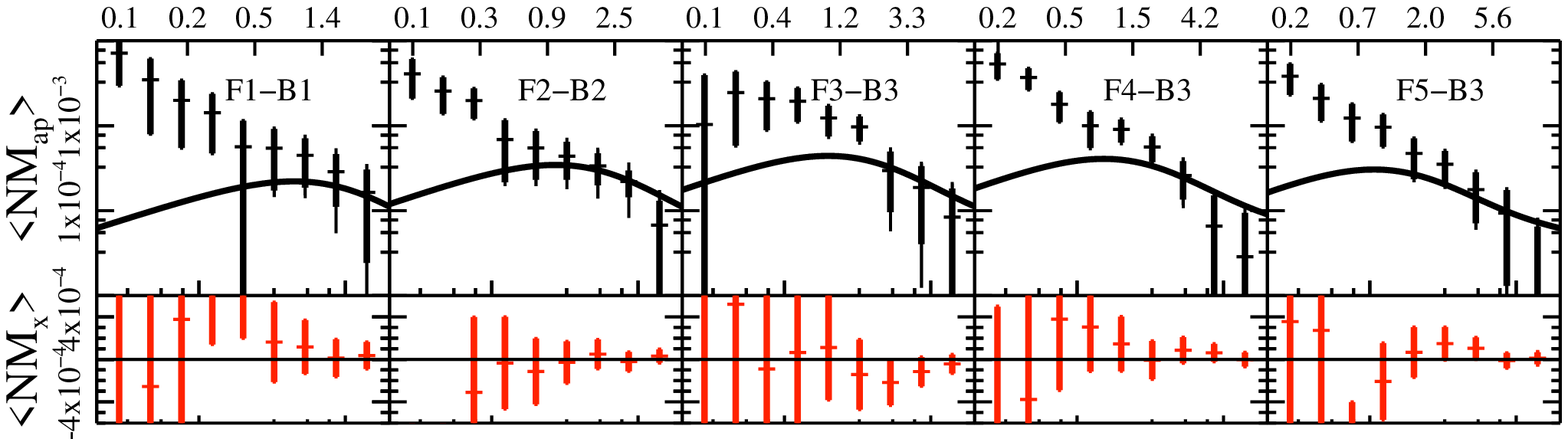}\\
        \includegraphics[width=0.95\linewidth]{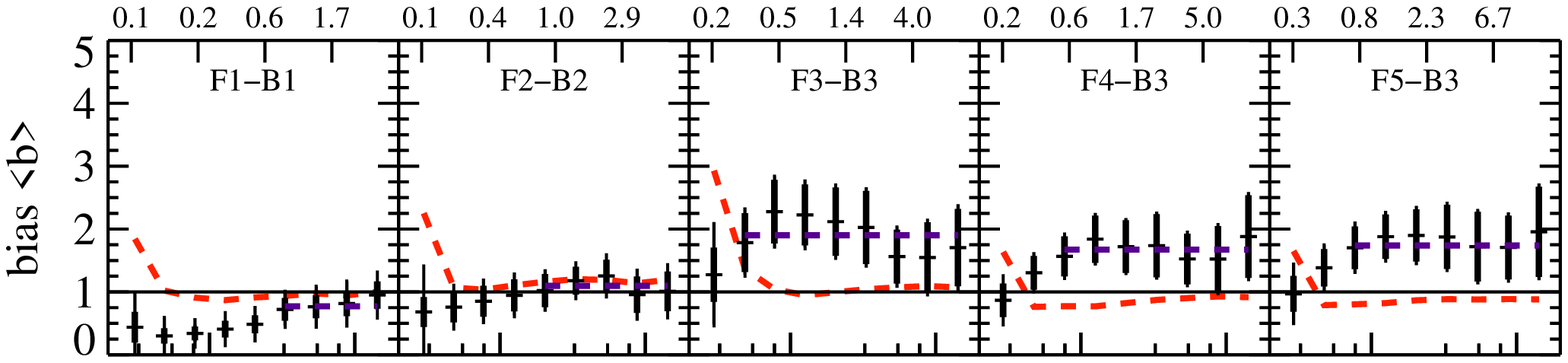}\\
        \includegraphics[width=0.95\linewidth]{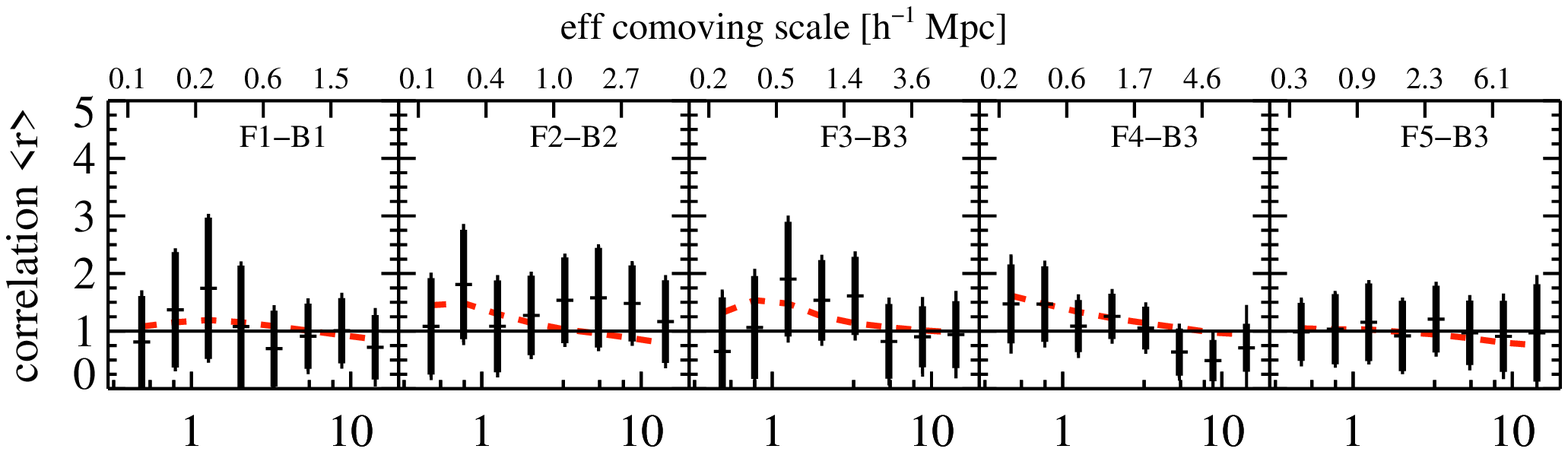}\\
    \end{tabular}
    \caption{Same as Figure~\ref{fig:b_m9}, but for galaxies in the
    stellar mass range $10^{10} < M_* < 10^{11}$.}
    \label{fig:b_m11}

\end{figure*}

\begin{figure*}
    \centering
    \begin{tabular}{cc}
        \includegraphics[width=0.45\textwidth]{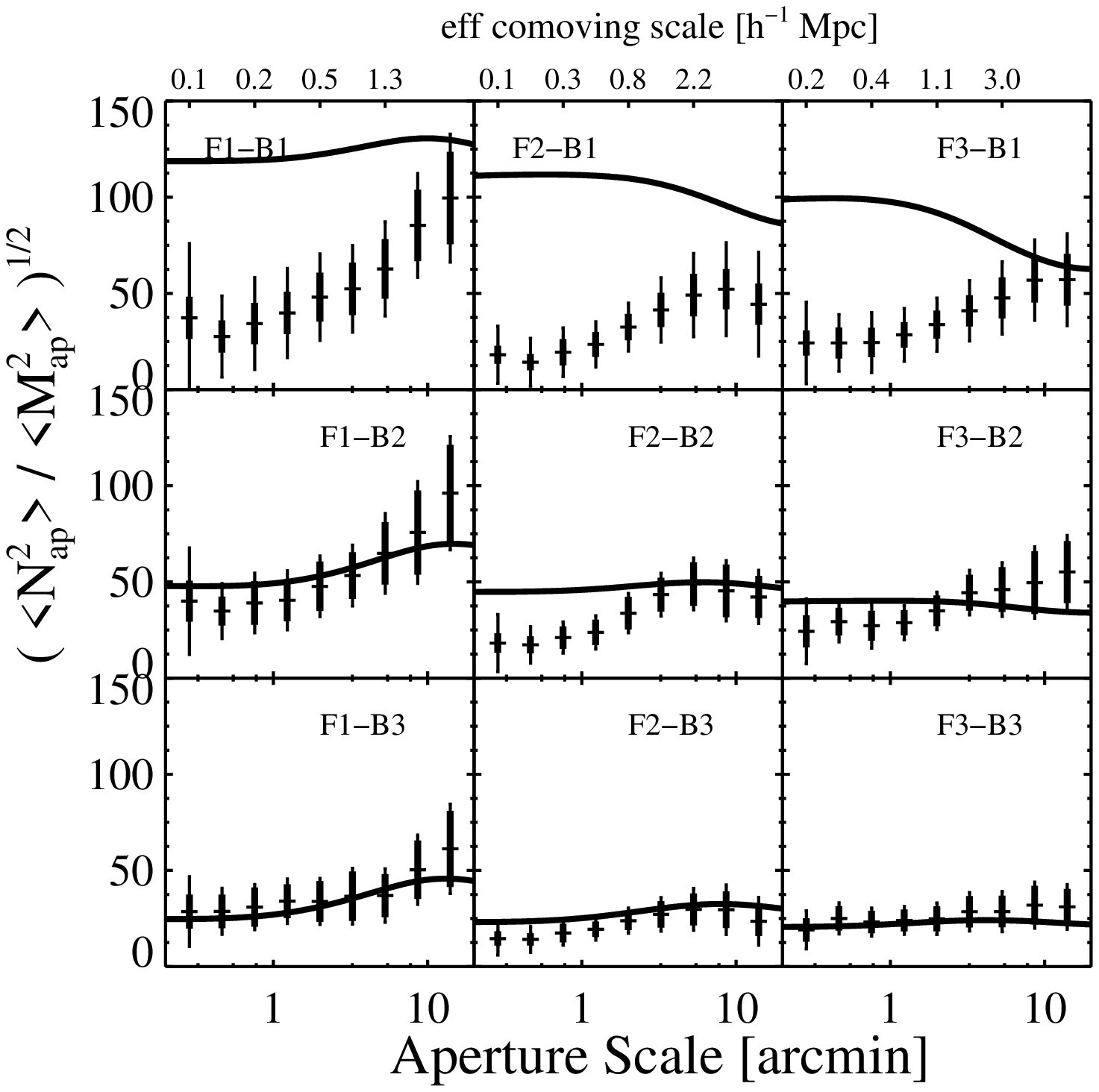} &
        \includegraphics[width=0.45\textwidth]{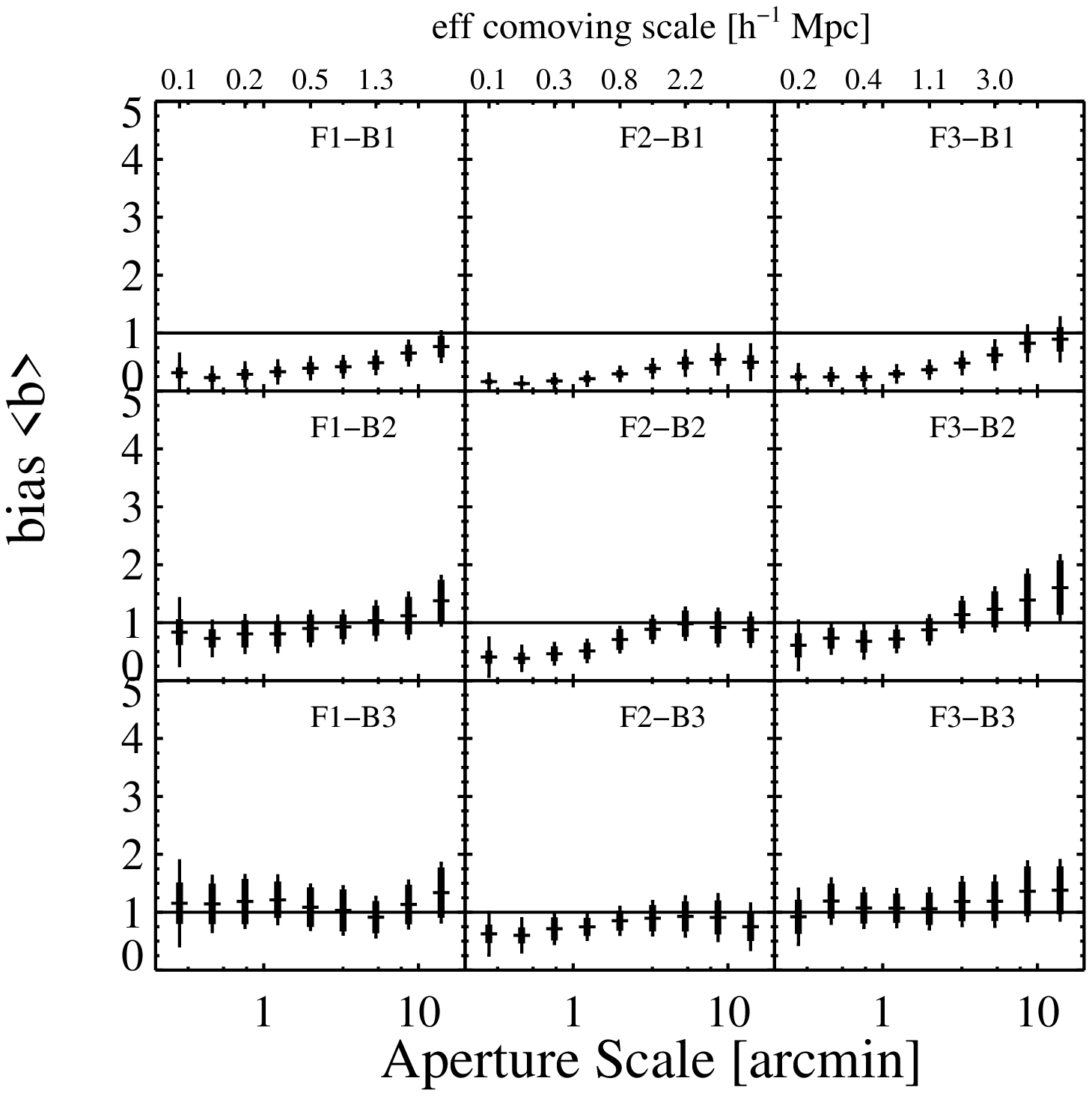}
    \end{tabular}
    \caption{The effects of a mismatch between foreground (F1, F2, F3) and
    background (B1, B2, B3) redshift bins for the stellar mass
    selected galaxy sample $10^9 < M_* < 10^{10}$. {\bf Left panel~|} The bias
    parameter $\b$ not calibrated by function $f1$. The thick line is
    the calibration signal $1/f_1$ computed assuming $b=1$. {\bf Right
    panel~|} The bias parameter $\b$ calibrated by function $f1$. }

    \label{fig:mismatch}
\end{figure*}

\end{document}